\documentclass[12pt,preprint,number,sort&compress,lontitle,times]{elsarticle}
\usepackage{natbib}
\usepackage{graphicx}
\usepackage{epsfig}
\usepackage{lineno}
\usepackage{multirow}
\usepackage{fourier}
\usepackage{color}
\usepackage{amsmath}

\makeatletter
\newcommand*{\rom}[1]{\expandafter\@slowromancap\romannumeral #1@}

\begin{document}

\begin{frontmatter}
\title{Searches for Large-Scale Anisotropies of Cosmic Rays~:\\
Harmonic Analysis and Shuffling Technique}
\author{O. Deligny, F. Salamida\\
Institut de Physique Nucl\'eaire\\
CNRS/IN2P3 \& Universit\'e Paris Sud, Orsay, France}

\begin{abstract}
The measurement of large scale anisotropies in cosmic ray arrival directions is generally 
performed through harmonic analyses of the right ascension distribution as a function
of energy. These measurements are challenging due to the small expected anisotropies
and meanwhile the relatively large modulations of observed counting rates due to 
experimental effects. In this paper, we present a procedure based on the shuffling technique
to carry out these measurements, applicable to any cosmic ray detector without any 
additional corrections for the observed counting rates. 
\end{abstract}
\end{frontmatter}

\section{Introduction}

In investigating the origin of cosmic rays, the measurement of anisotropy in the arrival 
directions is an important tool, complementary to the energy spectrum and mass composition.
From the observational point of view indeed, the study of the cosmic ray anisotropy with energy
is closely connected to the problem of cosmic ray propagation and sources. Due to the 
scattering of cosmic rays in the Galactic magnetic field, the anisotropy imprinted in arrival
directions is mainly expected at large scales up to the highest energies. The experimental 
study of large scale anisotropies is thus fundamental for cosmic ray physics, though it is
challenging. 

At energies greater than $10^{13}$~eV, cosmic ray measurements are indirect due to the
low primary intensity, and are usually performed through extensive air showers with surface
array detectors, Cherenkov telescopes, or fluorescence telescopes. Surface array detectors
with high duty cycle operate almost uniformly with respect to sidereal time thanks to the rotation 
of the Earth. In contrast, the low duty cycle of Cherenkov and fluorescence telescopes, operating
only on dark nights, propagates into large variations of exposure with sidereal time. These 
variations, almost always larger than the expected anisotropies, induce artificial modulations 
of the cosmic ray intensity in local sidereal time, and thus make delicate the measurements of 
harmonics in the right ascension distribution - in particular for the first harmonic. 

The most commonly used technique to search for large scale anisotropies is the analysis in right 
ascension only, through harmonic analysis (referred to as Rayleigh formalism) of the event counting 
rate~\cite{Linsley1975}. The technique itself is rather simple, but the greatest difficulties are 
in the estimation of the \textit{directional exposure} of the experiment. For any energy range, the 
directional exposure $\tilde{\omega}(\delta,\alpha)$ provides the effective time-integrated collecting 
area for a flux from each declination $\delta$ and right ascension $\alpha$~:
\begin{equation}
\label{eqn:exposure2d}
\tilde{\omega}(\delta,\alpha)=A~\int \mathrm{d}T~F(\theta(T-\alpha,\delta),\varphi(T-\alpha,\delta),T).
\end{equation}
Here, $T$ stands for the local sidereal time, and $A$ stands for the total collecting area of the 
experiment. The function $F(\theta,\varphi,T)$ encompasses geometrical aperture and detection 
efficiency effects under incidence zenith angle $\theta$ and azimuth angle $\varphi$. The explicit 
dependence of $F$ on local sidereal time $T$ is due to the unavoidable variations of the on-time 
of the detectors. In addition, extensive air showers properties are affected by meteorological 
modulations of the air density (through variations of the Moli\`ere radius) and of the pressure 
(due to the absorption of the electromagnetic component)~\cite{Lloyd1982,Auger2009}. This can 
produce a seasonal variation of the diurnal counting rate, which then results in an artificial modulation
of the intensity in local sidereal time~\cite{Farley1954}. The accurate control of all these effects is 
challenging in the case of surface array detectors, and generally out of reach in cases of
Cherenkov and fluorescence telescopes. Note that, when searching for large-scale anisotropies 
in right ascension only, the directional exposure $\omega(\alpha)$ integrated in declination is in fact 
the quantity of interest~:
\begin{equation}
\label{eqn:exposure}
\omega(\alpha)=\int \mathrm{d}\delta\cos{\delta}~\tilde{\omega}(\delta,\alpha).
\end{equation}
Throughout the paper, the use of 'directional exposure' will mainly refer to as $\omega(\alpha)$ - except
in sections~\ref{anisotropiccases} and~\ref{note} where explicit references to $\tilde{\omega}(\delta,\alpha)$
will be needed. In addition, the relative directional exposure $\omega_r(\alpha)$ will be also useful, 
defined as $\omega_r(\alpha)=\omega(\alpha)/\epsilon$, with $\epsilon$ the total exposure divided by $2\pi$. 

In contrast to large scale anisotropy searches, point-like source searches can be carried out 
by overcoming the explicit estimation of the $F$ function through the use of the 
\textit{shuffling method}~\cite{Cassiday1989}. This method only makes use of the 
\textit{observed} data set for determining the number
of background events in any direction of the sky, through the generation of simulation data sets
in which the actual event times are randomly associated with the actual local angles. In this
way, the counting rate variations are naturally accounted for in each simulation data set, because
all background events have the same time distribution as real events. In addition, preserving 
the local angle distribution as observed in the actual data set guarantees a proper modelling
of the detection efficiency in a total empirical manner. However, since events from eventual 
excesses are used to estimate the background level, the background estimate is necessarily 
overestimated compared to the true background~\cite{Cassiday1989,Alexandreas1993}. 
While this effect is negligible when searching for point sources, it is expected to be important 
when searching for diffuse excesses and in particular for large scale 
patterns~\cite{Rouille2007,Santos2008}. 

This paper is dedicated to explore in a comprehensive way the performances of the shuffling 
method when searching for large scale anisotropies in right ascension. Compared to previous 
studies, a new interpretation of the directional exposure function as estimated when applying the 
shuffling method is given, together with a complete procedure for interpreting the derived 
anisotropy amplitudes and for converting them into the corresponding anisotropy components 
in the equatorial plane. To this aim, the general formalism of harmonic analysis is first presented 
in section~\ref{sec:harmonicanalysis}, with a special attention given to the recovering of the
harmonic coefficients in the case of large variations of the directional exposure of the experiment 
in right ascension. The principle of the shuffling technique and its application to large 
scale anisotropy searches are then presented in section~\ref{sec:principle}. It is shown that some 
anisotropy can be recovered while properly accounting for any spurious effect of experimental
origin. The performances of this technique are given in section~\ref{sec:performances}
before to conclude in section~\ref{conclusions}.

\section{Harmonic analysis in right ascension}
\label{sec:harmonicanalysis}

Harmonic analysis of the right ascension distribution of cosmic rays in different \linebreak
energy ranges is a powerful tool for picking up and for characterising any modulation 
in this coordinate. Any angular distribution, $\Phi(\alpha)$, can be decomposed 
in terms of a harmonic expansion~:
\begin{equation}
\label{eqn:phi}
\Phi(\alpha)=a_0+\sum_{n>0}~a_n^c~\cos{n\alpha}+\sum_{n>0}~a_n^s~\sin{n\alpha}.
\end{equation}
The customary recipe to extract each harmonic coefficient makes use of the orthogona-
lity of the trigonometric functions~:
\begin{eqnarray}
\label{eqn:an}
a_0&=&\frac{1}{2\pi}\int_0^{2\pi}\mathrm{d}\alpha~\Phi(\alpha), \nonumber \\
a_n^c&=&\frac{1}{\pi}\int_0^{2\pi}\mathrm{d}\alpha~\Phi(\alpha)~\cos{n\alpha}, \nonumber \\
a_n^s&=&\frac{1}{\pi}\int_0^{2\pi}\mathrm{d}\alpha~\Phi(\alpha)~\sin{n\alpha}. 
\end{eqnarray}
In this section, we remind how this standard formalism can be applied to any set of
arrival directions $\{\alpha_1,...,\alpha_N\}_{1\leq i\leq N}$ in the case of a purely uniform
or a slightly non-uniform directional exposure, and present how to proceed in the case of
a highly non-uniform directional exposure. Hereafter, we use an over-line to indicate
the \textit{estimator} of any quantity.

\subsection{Uniform directional exposure}
In the case of a purely uniform directional exposure, the Rayleigh formalism~\cite{Linsley1975} 
directly provides the amplitude of the different harmonics, the corresponding phase (\textit{i.e.} right
ascension of the maximum of intensity), and the probability of detecting a signal due to fluctuations of an 
isotropic distribution with an amplitude equal or larger than the observed one. The observed 
arrival direction distribution, $\overline{\Phi}(\alpha)$, is here modelled as a sum of Dirac functions 
over the circle, $\overline{\Phi}(\alpha)=\sum_i \delta(\alpha,\alpha_i)$, so that integrations in
equation~\ref{eqn:an} reduce to discrete sums~:
\begin{eqnarray}
\label{eqn:an1}
\overline{a}_n^c&=&\frac{2}{N} \sum_{1\leq i \leq N} \cos{n\alpha_i}, \nonumber \\
\overline{a}_n^s&=&\frac{2}{N} \sum_{1\leq i \leq N} \sin{n\alpha_i}. 
\end{eqnarray}
Here, the re-calibrated harmonic coefficients $a_n^c\equiv a_n^c/a_0$ and $a_n^s\equiv a_n^s/a_0$
are directly considered, as it is traditionally the case in measuring \textit{relative} anisotropies.

The statistical properties of the estimators $\{\overline{a}_n^c,\overline{a}_n^s\}$ can be derived
from the Poissonian nature of the sampling of $N$ points over the circle distributed according to
the underlying angular distribution $\Phi(\alpha)$. From Poisson statistics indeed, the first and 
second moments of $\overline{\Phi}(\alpha)$ averaged over a large number of realisations of
$N$ events read~:
\begin{eqnarray}
\label{eqn:momentsphi}
\left\langle \overline{\Phi}(\alpha)\right\rangle&=&\epsilon\Phi(\alpha), \nonumber \\
\left\langle \overline{\Phi}(\alpha)\overline{\Phi}(\alpha^\prime)\right\rangle&=&\epsilon^2\Phi(\alpha)\Phi(\alpha^\prime)+\epsilon\Phi(\alpha)\delta(\alpha,\alpha^\prime). 
\end{eqnarray}
Propagating these properties into the first and second moments of 
the estimators $\{\overline{a}_n^c,\overline{a}_n^s\}$ leads, on the one hand, to unbiased estimators~:
\begin{eqnarray}
\label{eqn:moments1a}
\left\langle \overline{a}_n^c\right\rangle&=&a_n^c, \nonumber \\
\left\langle \overline{a}_n^s\right\rangle&=&a_n^s,
\end{eqnarray}
and, on the other hand, to the following covariance matrix~:
\begin{eqnarray}
\label{eqn:moments2a}
\mathrm{cov}(\overline{a}_m^c,\overline{a}_n^c)&=& \frac{1}{\epsilon\pi^2 a_0^2}\int\mathrm{d}\alpha~\Phi(\alpha)~\cos{m\alpha}~\cos{n\alpha}, \nonumber \\
\mathrm{cov}(\overline{a}_m^s,\overline{a}_n^s)&=& \frac{1}{\epsilon\pi^2 a_0^2}\int\mathrm{d}\alpha~\Phi(\alpha)~\sin{m\alpha}~\sin{n\alpha}, \nonumber \\
\mathrm{cov}(\overline{a}_m^c,\overline{a}_n^s)&=& \frac{1}{\epsilon\pi^2 a_0^2}\int\mathrm{d}\alpha~\Phi(\alpha)~\cos{m\alpha}~\sin{n\alpha}.
\end{eqnarray}
In case of small anisotropies (\textit{i.e.} $|a_n^c/a_0|\ll 1$ and $|a_n^s/a_0|\ll 1$), and with $N$ a 
good estimator of $\epsilon a_0$, the previous expressions allow the derivation of the RMS of the estimators as~:
\begin{equation}
\label{eqn:rmsa}
\sigma_n^c(\overline{a}_n^c)=\sigma_n^s(\overline{a}_n^s)=\bigg(\frac{2}{N}\bigg)^{0.5}.
\end{equation}

For an isotropic realisation, $\overline{a}_n^c$ and $\overline{a}_n^s$ are random variables
whose joint p.d.f., $p_{A_n^c,A_n^s}$, can be factorised in the limit of large number of events in 
terms of two Gaussian distributions whose variances are thus $\sigma^2=2/N$. For 
any $n$, the joint p.d.f. of the estimated amplitude, 
$\overline{r}_n=(\overline{a}_n^{c2}+\overline{a}_n^{s2})^{1/2}$, and phase,
$\overline{\phi}_n=\arctan{(\overline{a}_n^s/\overline{a}_n^c)}$, is then obtained through the 
Jacobian transformation~:
\begin{eqnarray}
\label{eqn:jointpdf1}
p_{R_n,\Phi_n}(\overline{r}_n,\overline{\phi}_n)&=& \left|\frac{\partial(\overline{a}_n^c,\overline{a}_n^s)}{\partial(\overline{r}_n,\overline{\phi}_n)} \right|~p_{A_n^c,A_n^s}(\overline{a}_n^c(\overline{r}_n,\overline{\phi}_n),\overline{a}_n^s(\overline{r}_n),\overline{\phi}_n) \nonumber \\
&=& \frac{\overline{r}_n}{2\pi\sigma^2}~\exp{(-\overline{r}_n^2/2\sigma^2)}.
\end{eqnarray}
From this expression, it is straightforward to recover the Rayleigh distribution for the p.d.f. of the
amplitude, $p_{R_n}$, and the uniform distribution between 0 and $2\pi$ for the p.d.f. of the
phase, $p_{\Phi_n}$.

\subsection{Non-uniform directional exposure}
\label{almostuniform}
The Rayleigh formalism aforementioned can be applied off the shelf only in the case of
a purely uniform directional exposure. This ideal
condition is generally not fulfilled, so that formally, the integrations performed in 
equation~\ref{eqn:an} do not allow any longer a direct extraction of the harmonic 
coefficients of $\Phi(\alpha)$. We assume they allow the extraction of the product
$\omega(\alpha)\Phi(\alpha)$, and will come back on the conditions of validity of this 
assumption in sub-section~\ref{anisotropiccases}.

At the \textit{sidereal} time scale, the directional exposure of most observatories operating with
high duty cycle (\textit{e.g.} surface detector arrays) is however only moderately (or even slightly) 
non-uniform. A simple recipe to account for the variations of $\omega(\alpha)$
is then to transform the observed angular distribution $\overline{\Phi}(\alpha)$ into the one
that would have been observed with an uniform directional exposure~: 
$\overline{\Phi}(\alpha)/\omega(\alpha)$~\cite{Mollerach2005}. In that way, discrete summations
in equation~\ref{eqn:an1} are changed into~:
\begin{eqnarray}
\label{eqn:an2}
\overline{a}_n^c&=&\frac{2}{\tilde{N}} \sum_{1\leq i \leq N} \frac{\cos{n\alpha_i}}{\omega(\alpha_i)}, \nonumber \\
\overline{a}_n^s&=&\frac{2}{\tilde{N}} \sum_{1\leq i \leq N} \frac{\sin{n\alpha_i}}{\omega(\alpha_i)}, 
\end{eqnarray}
with $\tilde{N}=\sum_i~1/\omega(\alpha_i)$. 

The statistical properties of these estimators can be estimated in the same way as in the previous
sub-section. The starting point is to consider the first and second moments of $\overline{\Phi}(\alpha)/\omega(\alpha)$~:
\begin{eqnarray}
\label{eqn:momentsphi2}
\left\langle \frac{\overline{\Phi}(\alpha)}{\omega(\alpha)} \right\rangle&=&\Phi(\alpha), \nonumber \\
\left\langle \frac{\overline{\Phi}(\alpha)}{\omega(\alpha)} \frac{\overline{\Phi}(\alpha^\prime)}{\omega(\alpha^\prime)} \right\rangle&=&\Phi(\alpha)\Phi(\alpha^\prime)+\frac{\Phi(\alpha)}{\omega(\alpha)}\delta(\alpha,\alpha^\prime). 
\end{eqnarray}
In case of small anisotropies, this leads to the following expressions for the RMS of the estimators~:
\begin{eqnarray}
\label{eqn:resolution1}
\sigma_n^c&=& \bigg(\frac{2}{N}\frac{1}{\pi}\int_0^{2\pi}\frac{\mathrm{d}\alpha}{\omega_r(\alpha)} \cos^2{n\alpha}\bigg)^{0.5}\nonumber, \\
\sigma_n^s&=& \bigg(\frac{2}{N}\frac{1}{\pi}\int_0^{2\pi}\frac{\mathrm{d}\alpha}{\omega_r(\alpha)} \sin^2{n\alpha}\bigg)^{0.5}.
\end{eqnarray}

The joint p.d.f. $p_{A_n^c,A_n^s}$ under isotropy is now the product of two Gaussian with two
distinct parameters $\sigma_n^c$ and $\sigma_n^s$. After the Jacobian transformation to get at the joint
p.d.f. $p_{R_n,\Phi_n}$, the integration over $\overline{\phi}_n$ yields the $p_{R_n}$ p.d.f.~:
\begin{equation}
\label{eqn:pdf-r}
p_{R_n}(\overline{r}_n)=\frac{\overline{r}_n}{\sigma_n^c\sigma_n^s}~\exp{\bigg(-\frac{{\sigma_n^c}^2+{\sigma_n^s}^2}{4{\sigma_n^c}^2{\sigma_n^s}^2}\overline{r}_n^2\bigg)}~I_0\bigg(\frac{{\sigma_n^c}^2-{\sigma_n^s}^2}{4{\sigma_n^c}^2{\sigma_n^s}^2}\overline{r}_n^2\bigg),
\end{equation}
where $I_0(\cdot)$ is the modified Bessel function of first kind with order 0. This p.d.f. can be 
viewed as a generalisation of the Rayleigh distribution in the case of non-equal resolutions
for the underlying Gaussian variables $\overline{a}_n^c$ and $\overline{a}_n^s$.
The probability of detecting a signal due to fluctuations of an isotropic distribution with an 
amplitude equal or larger than the observed one can be obtained by integrating numerically
$p_{R_n}$ above the observed value $\overline{r}_n$. On the other hand, the integration over 
$\overline{r}_n$ yields the $p_{\Phi_n}$ p.d.f.~:
\begin{equation}
\label{eqn:pdf-phi}
p_{\Phi_n}(\overline{\phi}_n)=\frac{\sigma_n^c\sigma_n^s}{2\pi}\frac{1}{{\sigma_n^c}^2\sin^2{\overline{\phi}_n}+{\sigma_n^s}^2\cos^2{\overline{\phi}_n}}.
\end{equation}
There are thus preferential directions for the reconstructed phases, fixed by the ratio between 
${\sigma_n^c}^2$ and ${\sigma_n^s}^2$ - independently of the number of events.  

This formalism is the relevant one in most of practical cases. It is to be noted, however, that variations 
of few percents in the directional exposure $\omega$ have no real numerical impact in 
equations~\ref{eqn:resolution1} (and so in equations~\ref{eqn:pdf-r} and~\ref{eqn:pdf-phi}), so that
in such cases, the p.d.f. of the amplitude can be approximated to a high level by the Rayleigh distribution, 
while the one of the phase by a uniform distribution.

\subsection{Highly non-uniform directional exposure}
\label{nonuniform}
Though the directional exposure of most observatories is generally only moderately non-uniform 
at the sidereal time scale, it is often useful to study other time scales for checking purposes. 
In such cases, the variation of the directional exposure can be large. At the solar time scale
for instance, and for observatories operating only during nights (\textit{e.g.} fluorescence
telescopes, Cherenkov telescopes), the directional exposure can be very small in the directions 
around the Sun, and even \textit{cancel}. Thus, it may be worth generalising the previous formalism to 
the case of \textit{any} directional exposure. 

For any given exposure function $\omega$, the harmonic expansion of the \textit{observed} 
angular distribution $\overline{\Phi}(\alpha)$ leads to the following linear system between 
the observed harmonic coefficients $\{b_n^c,b_n^s\}$ and the underlying $\{a_n^c,a_n^s\}$ 
ones~:
\begin{equation}
\label{eqn:b0}
b_0=\sum_{m\geq0}~\frac{a_m^c}{2\pi}\int_0^{2\pi}\mathrm{d}\alpha~\omega(\alpha)~\cos{m\alpha}
+\sum_{m>0}~\frac{a_m^s}{2\pi}\int_0^{2\pi}\mathrm{d}\alpha~\omega(\alpha)~\sin{m\alpha} \nonumber,
\end{equation}
\begin{equation}
\label{eqn:bodd}
b_n^c=\sum_{m\geq0}~\frac{a_m^c}{\pi}\int_0^{2\pi}\mathrm{d}\alpha~\omega(\alpha)~\cos{n\alpha}~\cos{m\alpha}
+\sum_{m>0}~\frac{a_m^s}{\pi}\int_0^{2\pi}\mathrm{d}\alpha~\omega(\alpha)~\cos{n\alpha}~\sin{m\alpha} \nonumber,
\end{equation}
\begin{equation}
\label{eqn:beven}
b_n^s=\sum_{m\geq0}~\frac{a_m^c}{\pi}\int_0^{2\pi}\mathrm{d}\alpha~\omega(\alpha)~\sin{n\alpha}~\cos{m\alpha}
+\sum_{m>0}~\frac{a_m^s}{\pi}\int_0^{2\pi}\mathrm{d}\alpha~\omega(\alpha)~\sin{n\alpha}~\sin{m\alpha}.
\end{equation}
Formally, the $\{a_n^c,a_n^s\}$ coefficients appear related to the $\{b_n^c,b_n^s\}$ ones through a 
convolution such that, in simplified notations, $b=K\cdot a$. The matrix $K$, entirely determined by 
the directional exposure, reads~:
\begin{equation}
\label{eqn:matrix}
K=\begin{bmatrix}
\vspace{0.3cm} \displaystyle\frac{1}{2\pi} \int \mathrm{d}\alpha~\omega(\alpha) & \displaystyle\frac{1}{2\pi} \int \mathrm{d}\alpha~\omega(\alpha)~\cos{\alpha} & \displaystyle\frac{1}{2\pi} \int \mathrm{d}\alpha~\omega(\alpha)~\sin{\alpha} & ... \\
\vspace{0.3cm} \displaystyle\frac{1}{\pi} \int \mathrm{d}\alpha~\omega(\alpha)~\cos{\alpha}&  \displaystyle\frac{1}{\pi} \int \mathrm{d}\alpha~\omega(\alpha)~\cos^2{\alpha} &  \displaystyle\frac{1}{\pi} \int \mathrm{d}\alpha~\omega(\alpha)~\cos{\alpha}~\sin{\alpha} & ...\\
\vspace{0.3cm} \displaystyle\frac{1}{\pi} \int \mathrm{d}\alpha~\omega(\alpha)~\sin{\alpha}& \displaystyle\frac{1}{\pi} \int \mathrm{d}\alpha~\omega(\alpha)~\cos{\alpha}~\sin{\alpha} & \displaystyle\frac{1}{\pi} \int \mathrm{d}\alpha~\omega(\alpha)~\sin^2{\alpha} & ... \\
...&...&...&...
\end{bmatrix}
\end{equation}
Meanwhile, the observed angular distribution $\overline{\Phi}(\alpha)$ can provide a direct estimation
of the $\{b_n^c,b_n^s\}$ coefficients through $\overline{b}_0=N/2\pi$ and~:
\begin{equation}
\label{eqn:bodd2}
\overline{b}_n^c=\frac{1}{\pi}~\sum_{1\leq i \leq N} \cos{n\alpha_i} \nonumber,
\end{equation}
\begin{equation}
\label{eqn:beven2}
\overline{b}_n^s=\frac{1}{\pi}~\sum_{1\leq i \leq N} \sin{n\alpha_i}.
\end{equation}
Then, if the angular distribution $\Phi(\alpha)$ has no higher harmonic moment than $n_{\mathrm{max}}$, the 
first $\{b_n^c,b_n^s\}$ coefficients with $n\leq n_{\mathrm{max}}$ are related to the non-vanishing $\{a_n^c,a_n^s\}$ 
by the square matrix $K_{n_{\mathrm{max}}}$ \textit{truncated} to $n_{\mathrm{max}}$. Inverting this truncated
matrix allows us to recover the underlying $\{a_n^c,a_n^s\}$ coefficients~: 
\begin{equation}
\label{eqn:recover}
\overline{a}=K^{-1}_{n_{\mathrm{max}}}\cdot \overline{b}.
\end{equation}

Relative to $b_0$, the resolution on each recovered $\{b_n^c,b_n^s\}$ coefficients is simply
$\sqrt{2/N}$. In case of small anisotropy amplitudes, the resolution on each recovered $\{a_n^c,a_n^s\}$ 
coefficients relative to $a_0$ can be estimated from the propagation of uncertainties, in the same way as
in previous sub-sections. This leads to~:
\begin{eqnarray}
\label{eqn:resolution}
\sigma_n^c&=& \bigg([^tK^{-1}_{n_{\mathrm{max}}}]_{2n,2n}/(\pi\overline{a}_0)\bigg)^{0.5}\nonumber, \\
\sigma_n^s&=& \bigg([^tK^{-1}_{n_{\mathrm{max}}}]_{2n+1,2n+1}/(\pi\overline{a}_0)\bigg)^{0.5}.
\end{eqnarray}
The p.d.f. $p_{R_n}$ for the amplitude and $p_{\Phi_n}$ for the phase under isotropy are exactly the same 
as in the previous sub-section - \textit{cf} equations~\ref{eqn:pdf-r} and~\ref{eqn:pdf-phi} - with these resolution
parameters.

It is worth noting that if the directional exposure $\omega$ does \textit{not} cancel, this formalism 
considered for $n_{\mathrm{max}}\rightarrow \infty$ is actually \textit{identical} to the one presented in previous 
sub-section~\ref{almostuniform}. For $n_{\mathrm{max}}\rightarrow \infty$ indeed, the completeness 
relation of the harmonic functions between 0 and $2\pi$ allows us to express the $K^{-1}_\infty$ matrix as~:
\begin{equation}
\label{eqn:matrixinv}
K^{-1}_\infty=\begin{bmatrix}
\vspace{0.3cm} \displaystyle\frac{1}{2\pi} \int \frac{\mathrm{d}\alpha}{\omega(\alpha)} & \displaystyle\frac{1}{2\pi} \int \frac{\mathrm{d}\alpha}{\omega(\alpha)}~\cos{\alpha} & \displaystyle\frac{1}{2\pi} \int \frac{\mathrm{d}\alpha}{\omega(\alpha)}~\sin{\alpha} & ... \\
\vspace{0.3cm} \displaystyle\frac{1}{\pi} \int \frac{\mathrm{d}\alpha}{\omega(\alpha)}~\cos{\alpha}&  \displaystyle\frac{1}{\pi} \int \frac{\mathrm{d}\alpha}{\omega(\alpha)}~\cos^2{\alpha} &  \displaystyle\frac{1}{\pi} \int \frac{\mathrm{d}\alpha}{\omega(\alpha)}~\cos{\alpha}~\sin{\alpha} & ...\\
\vspace{0.3cm} \displaystyle\frac{1}{\pi} \int \frac{\mathrm{d}\alpha}{\omega(\alpha)}~\sin{\alpha}& \displaystyle\frac{1}{\pi} \int \frac{\mathrm{d}\alpha}{\omega(\alpha)}~\cos{\alpha}~\sin{\alpha} & \displaystyle\frac{1}{\pi} \int \frac{\mathrm{d}\alpha}{\omega(\alpha)}~\sin^2{\alpha} & ... \\
...&...&...&...
\end{bmatrix}
\end{equation}
It is then straightforward to see that the relations
\begin{eqnarray}
\label{eqn:identity}
\sum_{m=2p} \left[ K^{-1}_\infty \right]_{nm} \cos{\frac{m\alpha}{2}}+\sum_{m=2p+1} \left[ K^{-1}_\infty \right]_{nm} \sin{\frac{(m+1)\alpha}{2}}=\left\{\begin{array}{ll}
\displaystyle\frac{\cos{n\alpha}}{\omega(\alpha)} \mathrm{if}~n~\mathrm{even}\\
\displaystyle\frac{\sin{n\alpha}}{\omega(\alpha)} \mathrm{if}~n~\mathrm{odd}\end{array}\right.
\end{eqnarray}
hold. Inserting these relationships into equation~\ref{eqn:recover}, it can be seen that the estimators
defined in equation~\ref{eqn:an2} and~\ref{eqn:recover} are identical.

\subsection{Anisotropic cases}
\label{anisotropiccases}

\subsubsection{Uniform exposure}
Up to now, the amplitude p.d.f. $p_{R_n}$ and phase p.d.f. $p_{\Phi_n}$ have been derived in the case of an underlying
\textit{isotropic} distribution. It will be useful in the following to know as well the expected distributions in the case of an 
anisotropic distribution characterised by non-zero $\{a_n^c=\nu_n\cos{\psi_n},a_n^s=\nu_n\sin{\psi_n}\}$ harmonic 
coefficients. The principle for deriving the $p_{N_n}$ and $p_{\Psi_n}$ distributions is the same as in the isotropic case~:
it consists first in deriving the joint p.d.f. $p_{N_n,\Psi_n}$, and then to marginalise $p_{N_n,\Psi_n}(\nu_n,\psi_n)$ over 
$\overline{\phi}_n$ and $\overline{r}_n$ for obtaining the $p_{N_n}$ and $p_{\Psi_n}$ distributions respectively. The 
joint p.d.f. $p_{A_n^c,A_n^s}$ is now expressed in terms of two Gaussian distributions centered on 
$\{a_n^c=\nu_n\cos{\psi_n},a_n^s=\nu_n\sin{\psi_n}\}$, so that the Jacobian transformation for obtaining $p_{N_n,\Psi_n}$ 
reads now~:
\begin{equation}
\label{eqn:jointpdf2}
p_{N_n,\Psi_n}(\overline{r}_n,\overline{\phi}_n;s_n,\psi_n)=\left|\frac{\partial(\overline{a}_n^c,\overline{a}_n^s)}{\partial(\overline{r}_n,\overline{\phi}_n)} \right|~p_{A_n^c,A_n^s}(\overline{a}_n^c(\overline{r}_n,\overline{\phi}_n)-a_n^c(\nu_n,\psi_n),\overline{a}_n^s(\overline{r}_n),\overline{\phi}_n-a_n^s(\nu_n,\psi_n)).
\end{equation}

For convenience, we drop the indexes $n$ hereafter. In the case of a uniform directional exposure, the integration of 
equation~\ref{eqn:jointpdf2} over $\overline{\phi}$ yields to a non-centered Rayleigh distribution~\cite{Linsley1975}~:
\begin{equation}
\label{eqn:pdf-rice}
p_{N}(\overline{r};\nu)=\frac{\overline{r}}{\sigma^2} \exp{\bigg(-\frac{\overline{r}^2+\nu^2}{2\sigma^2}\bigg)}I_0\bigg(\frac{\overline{r}\nu}{\sigma^2}\bigg),
\end{equation}
while the integration over $\overline{r}$ yields to~\cite{Linsley1975}~:
\begin{equation}
\label{eqn:pdf-psi}
p_{\Psi}(\overline{\phi};\nu,\psi)=\frac{1}{2\pi}\exp{\bigg(-\frac{\nu^2}{2\sigma^2}}\bigg)+\frac{\nu\cos{(\overline{\phi}-\psi)}}{2\sqrt{2\pi}\sigma}\bigg(1+\mathrm{erf}{\bigg(\frac{\nu\cos{(\overline{\phi}-\psi)}}{\sqrt{2}\sigma}\bigg)}\bigg)\exp{\bigg(-\frac{\nu^2\sin^2{(\overline{\phi}-\psi)}}{2\sigma^2}\bigg)}.
\end{equation}
In the case of an angular distribution on the sphere such that the flux of cosmic rays $\tilde{\Phi}(\delta,\alpha)$ 
can be expressed in terms of a monopole and a dipole characterised by a vector $\mathbf{d}=\{d\cos{\delta_d}\cos{\alpha_d},d\cos{\delta_d}\sin{\alpha_d},d\sin{\delta_d}\}$ 
in equatorial coordinates, it is interesting to relate the dipole parameters $\{d,\delta_d,\alpha_d\}$ to the 
ones as derived from the Rayleigh analysis $\{\nu,\psi\}$. This can be done by inserting in equations~\ref{eqn:an} 
the expression of $\Phi(\alpha)$ in terms of the flux $\tilde{\Phi}(\delta,\alpha)$~:
\begin{equation}
\label{eqn:flux2d}
\Phi(\alpha)=\int \mathrm{d}\delta~\cos{\delta}~\tilde{\omega}(\delta)\tilde{\Phi}(\delta,\alpha),
\end{equation}
where the function $\tilde{\omega}$ is the two-dimensional directional exposure defined in equation~\ref{eqn:exposure2d}, 
which is here independent of the right-ascension. Plugging this expression into equations~\ref{eqn:an} leads to~\cite{Aublin2005}~:
\begin{eqnarray}
\label{eqn:nu}
\nu&=&\frac{d\cos{\delta_d}\left\langle\cos{\delta}\right\rangle}{1+d\sin{\delta_d}\left\langle\sin{\delta}\right\rangle} \nonumber,\\
\psi&=&\alpha_d.
\end{eqnarray}
The first harmonic amplitude $\nu$ depends on the declination of the dipole in such a way that it vanishes 
for $\delta_d=\pm \pi/2$ while it is the largest when the dipole is oriented in the equatorial plane. In this 
later case, the first harmonic amplitude simply becomes $\nu=\left<\cos{\delta}\right>d_\perp$ (with 
$d_\perp=d\cos{\delta_d}$) and the sensitivity of any observatory to $d_\perp$ depends then on 
$\left<\cos{\delta}\right>$, which is a function of the Earth latitude $\ell$ of the experiment and of its 
detection efficiency in the zenithal range considered. 

\subsubsection{Non-uniform exposure}
\begin{figure}[!t]
  \centering					 
 \includegraphics[width=7.5cm]{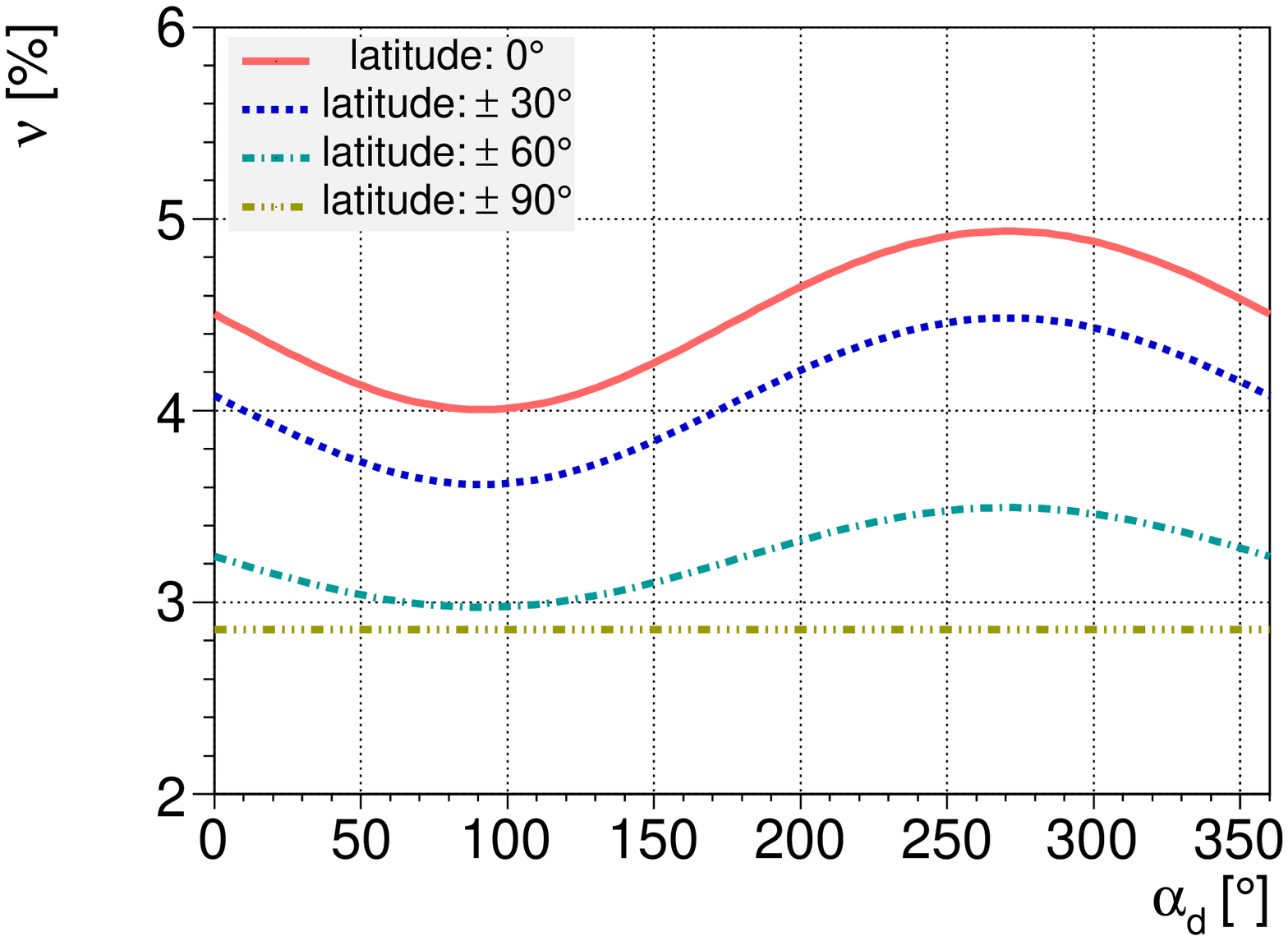}
 \includegraphics[width=7.5cm]{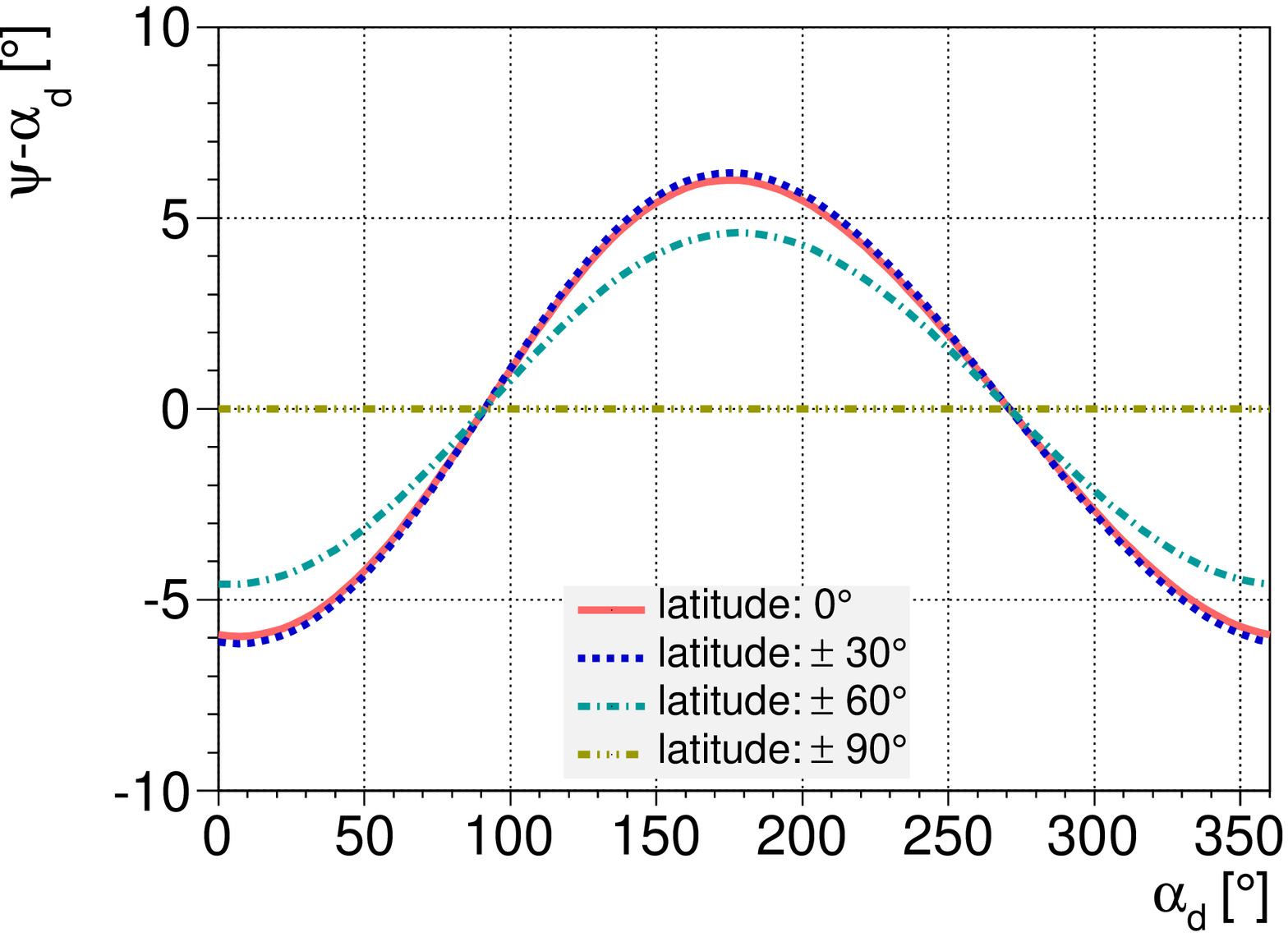}
  \caption{\small{Variations of the $\nu$ (left) and $\psi$ (right) parameters as a function of the dipole phase $\alpha_d$ 
  for a dipole amplitude $d=5\%$ \textit{in the case of a non-uniform directional exposure}, for different latitudes of virtual 
  observatories.}}
  \label{fig:nupsi}
\end{figure}
Turning now to the case of a non-uniform directional exposure in right ascension, and considering a general
cosmic ray flux $\tilde{\Phi}$ depending on both $\delta$ and $\alpha$, it turns out from equation~\ref{eqn:flux2d} 
that the assumption made in sub-sections~\ref{almostuniform} and~\ref{nonuniform}, namely that the observed 
distribution in right ascension is the product between the directional exposure $\omega(\alpha)$ and the function 
$\Phi(\alpha)$, is in fact not fulfilled if the function $\tilde{\omega}(\delta,\alpha)$ cannot be factorised - and there
is no reason that this function could be factorised in practical cases. Formally, this implies that the recovered 
harmonic coefficients are necessarily biased since genuine anisotropy effects cannot be fully decoupled from exposure 
effects. Considering for simplicity the estimates defined in sub-section~\ref{almostuniform} only, this can be
easily seen from the formal expression of the estimators~:
\begin{eqnarray}
\label{eqn:nu2}
\tilde{N}&=&\int\mathrm{d}\Omega~\frac{\tilde{\omega}(\delta,\alpha)}{\omega(\alpha)}~\tilde{\Phi}(\delta,\alpha), \nonumber \\
a_n^c&=&\frac{2}{\tilde{N}}\int\mathrm{d}\Omega~\frac{\tilde{\omega}(\delta,\alpha)}{\omega(\alpha)}~\tilde{\Phi}(\delta,\alpha)~\cos{n\alpha}, \nonumber \\
a_n^s&=&\frac{2}{\tilde{N}}\int\mathrm{d}\Omega~\frac{\tilde{\omega}(\delta,\alpha)}{\omega(\alpha)}~\tilde{\Phi}(\delta,\alpha)~\sin{n\alpha}, 
\end{eqnarray}
which, given the definitions of $\omega(\alpha)$ and $\Phi(\alpha)$, cannot be written in the same form as 
equations~\ref{eqn:an} (except if $\tilde{\Phi}$ depends only on $\alpha$, which is a very specific case).

We exemplify here that, 
however, the biases are not large even in a case of a relatively large variation of $\omega(\alpha)$. For $n=1$
for instance, the relationship between the dipole vector $\mathbf{d}$ parameters and the first harmonic
parameters $\{\nu,\psi\}$ can be calculated numerically from equations~\ref{eqn:nu2}. Some 
illustrations of this relation are shown in Figure~\ref{fig:nupsi} for $d=5\%,\delta_d=0^\circ$ and different
latitudes of virtual observatories fully efficient up to $60^\circ$ and with a variation of the 
detection efficiency with local sidereal time $T$ proportional to $1+0.3\cos{T}$. 
The time dependence of the detection efficiency translates into a right ascension dependence of the 
directional exposure, except for observatory latitudes located at the poles of the Earth where the whole 
range of right ascensions is in the field of view at any time, resulting in a uniform exposure. It can be seen that 
both $\nu$ and $\psi$ undergo now some variations with the dipole phase $\alpha_d$ for a wide range of latitudes,
except, as expected, for those located at the poles of the Earth where the distance between $\nu$ and $d$ is, 
however, the largest. This kind of behaviour will be important to consider for understanding the performances 
of the shuffling method in sub-section~\ref{anisotropy}.

For completeness, we provide also the amplitude and phase p.d.f. expected for non-zero $\nu$. Only semi-analytical 
expressions can be derived~:
\begin{eqnarray}
\label{eqn:pdfanis}
p_{N}(\overline{r};\nu,\psi)&=&\frac{\overline{r}}{2\pi\sigma^c\sigma^s}\exp{\bigg(-\frac{\nu^2\cos^2{\psi}}{2{\sigma^c}^2}-\frac{\nu^2\sin^2{\psi}}{2{\sigma^s}^2}\bigg)}\times\nonumber\\
&\displaystyle\int& \mathrm{d}\overline{\phi}~\exp{\bigg(-\frac{\overline{r}^2\cos^2{\overline{\phi}}}{2{\sigma^c}^2}-\frac{\overline{r}^2\sin^2{\overline{\phi}}}{2{\sigma^s}^2}+\frac{\overline{r}\nu\cos{\psi}\cos{\overline{\phi}}}{{\sigma^c}^2}+\frac{\overline{r}\nu\sin{\psi}\sin{\overline{\phi}}}{{\sigma^s}^2}\bigg)},\\
p_{\Psi}(\overline{\phi};\nu,\psi)&=&\frac{1}{2\pi\sigma^c\sigma^s}\exp{\bigg(-\frac{\nu^2\cos^2{\psi}}{2{\sigma^c}^2}-\frac{\nu^2\sin^2{\psi}}{2{\sigma^s}^2}\bigg)}\times\nonumber\\
&\displaystyle\int&\mathrm{d}\overline{r}~\overline{r}\exp{\bigg(-\frac{\overline{r}^2\cos^2{\overline{\phi}}}{2{\sigma^c}^2}-\frac{\overline{r}^2\sin^2{\overline{\phi}}}{2{\sigma^s}^2}+\frac{\overline{r}\nu\cos{\psi}\cos{\overline{\phi}}}{{\sigma^c}^2}+\frac{\overline{r}\nu\sin{\psi}\sin{\overline{\phi}}}{{\sigma^s}^2}\bigg)}.
\end{eqnarray}

\subsection{Note on the dipolar interpretation of the first harmonic coefficients}
\label{note}

The parameters $\nu$ and $\psi$ are often interpreted as indirect estimations of the dipole parameters $d$ and
$\alpha_d$, based on the relationships established in equations~\ref{eqn:nu} or~\ref{eqn:nu2}.
Actually, for a general angular distribution $\tilde{\Phi}(\delta,\alpha)$ with multipolar moments beyond the dipole, 
these simple relationships do not hold anymore and there is no simple way to relate the first harmonic coefficients
to the dipole parameters anymore. 

For convenience, this statement is demonstrated here only in the case of a uniform exposure function in right ascension. 
In its most general form, the flux on the sphere can be decomposed as an infinite sum of spherical harmonics 
$Y_{\ell m}(\delta,\alpha)$ with multipolar coefficients $a_{\ell m}$. Inserting this expansion into equation~\ref{eqn:flux2d},
the angular distribution $\Phi(\alpha)$ can be expressed as~:
\begin{equation}
\label{eqn:flux2dbis}
\Phi(\alpha)\propto\int \mathrm{d}\delta~\cos{\delta}~\tilde{\omega}(\delta)\bigg(1+\sum_{\ell\geq1}\sum_{m=-\ell}^{\ell} a_{\ell m}Y_{\ell m}(\delta,\alpha)\bigg).
\end{equation}
Using the fact that $Y_{\ell m}(\delta,\alpha)$ functions are proportional to the product of the associated Legendre 
polynomials $P_{\ell m}$ for $\sin{\delta}$ and of harmonic functions for $\alpha$, the explicit dependence of $\Phi(\alpha)$ 
in $\cos{\alpha}$ and $\sin{\alpha}$ can be written as~:
\begin{eqnarray}
\label{eqn:flux2dter}
\Phi(\alpha)&\propto& 1+\sum_{\ell\geq1} \sqrt{\frac{(2\ell+1)(\ell-1)!}{2\pi(\ell+1)!}} \left\langle P_{\ell 1}(\sin{\delta}) \right\rangle \bigg(a_{\ell 1}\cos{\alpha}+a_{\ell -1}\sin{\alpha}\bigg)+...,
\end{eqnarray}
with, as in equation~\ref{eqn:nu}, the shortcut notation 
\begin{equation}
\left\langle P_{\ell 1}(\sin{\delta}) \right\rangle=\frac{\displaystyle\int \mathrm{d}\delta~\cos{\delta}~\tilde{\omega}(\delta)P_{\ell 1}(\sin{\delta})}{\displaystyle\int \mathrm{d}\delta~\cos{\delta}~\tilde{\omega}(\delta)}.
\end{equation}
By identifying the first harmonic coefficients in equations~\ref{eqn:phi} and~\ref{eqn:flux2dter}, the general
relationship between $\{\nu,\psi\}$ and the $a_{\ell \pm1}$ coefficients is now~:
\begin{equation}
\label{eqn:nupsibis}
\nu=\frac{\bigg[\bigg(\sum_{\ell\geq1}\sqrt{\frac{(2\ell+1)(\ell-1)!}{2\pi(\ell+1)!}}\left\langle P_{\ell 1}(\sin{\delta}) \right\rangle a_{\ell 1}\bigg)^2+\bigg(\sum_{\ell\geq1}\sqrt{\frac{(2\ell+1)(\ell-1)!}{2\pi(\ell+1)!}}\left\langle P_{\ell 1}(\sin{\delta}) \right\rangle a_{\ell -1}\bigg)^2\bigg]^{0.5}}{1+\sum_{\ell\geq0}\sqrt{\frac{(2\ell+1)}{4\pi}}\left\langle P_{\ell 0}(\sin{\delta}) \right\rangle a_{\ell 0}},
\end{equation}
\begin{equation}
\label{eqn:nupsiter}
\tan{\psi}=\frac{\sum_{\ell\geq1}\sqrt{\frac{(2\ell+1)(\ell-1)!}{(\ell+1)!}}\left\langle P_{\ell 1}(\sin{\delta}) \right\rangle a_{\ell -1}}{\sum_{\ell\geq1}\sqrt{\frac{(2\ell+1)(\ell-1)!}{(\ell+1)!}}\left\langle P_{\ell 1}(\sin{\delta}) \right\rangle a_{\ell 1}}.
\end{equation}
For an exposure function $\tilde{\omega}$ covering mainly one of the hemisphere of the Earth only, it is thus permissible,
for instance, that some families of quadrupoles produce a non-zero first harmonic in right ascension, without any dipole pattern in
the underlying angular distribution $\tilde{\Phi}$. Hence, for interpreting any measurement of first harmonic in right ascension
in terms of dipole amplitude and phase, the additional assumption that the flux $\tilde{\Phi}$ is a pure dipole is needed\footnote{
Actually, the additional assumption can be formulated in a slightly weaker way~: the flux $\tilde{\Phi}$ does not contain any
non-zero $a_{\ell \pm1}$ coefficients except for $\ell=1$.}.

\section{Principle of the shuffling technique for large scale anisotropy searches}
\label{sec:principle}

\subsection{Principle of the shuffling technique}
\label{principle}
Under the hypothesis of isotropy, any shower detected with particular local coordinates
could have arrived with equal probability at any other time of a shower detection. The
shuffling technique simply exploits this property, in estimating the directional exposure 
by averaging the number of arrivals in a given target over a large number of simulation 
data sets~\cite{Cassiday1989}. Each simulation data set is obtained by preserving 
the actual set of arrival directions in local coordinates, and by randomly sampling 
times from the actual set of measured ones. 

Formally, the principle of this approach is to replace the time integration and the 
$\cos{\delta}$ integration of the
collecting area in equation~\ref{eqn:exposure} by the solid angle and time integrations 
of the observed angular distribution $\mathrm{d}N/\mathrm{d}\Omega$ in local angles 
$\Omega=(\theta,\varphi)$ pointing to the right ascension $\alpha$ at local sidereal 
time $T$ and weighted by the observed and normalised event time distribution 
$1/N\times\mathrm{d}N/\mathrm{d}T$, $N$ being the total number of events~:
\begin{equation}
\label{eqn:shuff1}
\omega_{sh}(\alpha)= \int \mathrm{d}\Omega \mathrm{d}T \frac{1}{N}~\frac{\mathrm{d}N(T)}{\mathrm{d}T}~\frac{\mathrm{d}N(\Omega)}{\mathrm{d}\Omega}~\delta(\alpha(\theta,\varphi, T)-\alpha).
\end{equation}
The argument in the Dirac function guarantees that the direction in celestial coordinates
considered throughout the integration of the local angles at local sidereal time $T$ 
corresponds to the right ascension $\alpha$~:
\begin{equation}
\label{eqn:shuff2}
\alpha(\theta,\varphi,T)=T+f(\varphi)~\arccos{\left( \frac{\cos{\theta}-\cos{\ell}\cos{\delta}}{\sin{\ell}\sin{\delta}}\right)},
\end{equation}
with $f(\varphi)=+1$ if $-\pi/2\leq\varphi\leq\pi/2$ and $f(\varphi)=-1$ otherwise for an
azimuth angle $\varphi$ defined relative to the East direction and measured counterclockwise.
Integrating over solid angle $\Omega=(\theta,\varphi)$ the angular distribution 
$\mathrm{d}N/\mathrm{d}\Omega$ results in giving to any celestial direction in 
the field of view available at any given local sidereal time $T$ an instantaneous exposure 
in proportion to the event rate in the corresponding direction in local angles. Through the 
integration over local sidereal time $T$ of the actual variation of the event rate 
$\mathrm{d}N/\mathrm{d}T$, the distortions of the cosmic ray intensity in celestial 
coordinates induced by experimental variations of the event rate are then automatically accounted 
for in the definition of $\omega_{sh}(\alpha)$. In contrast, the modulations in celestial coordinates 
induced by eventual anisotropies are only \textit{partially} washed out, because for any given 
local sidereal time $T$, the event time distribution $\mathrm{d}N/\mathrm{d}T$
is sensitive to the global intensity of cosmic rays but not to the underlying structure in right 
ascension. Hence, the use of equation~\ref{eqn:shuff1} provides a relevant estimate of the expected 
background in searching for large scale anisotropies, though with reduced sensitivity.

The main advantage in adopting equation~\ref{eqn:shuff1} as the directional exposure
instead of the actual one relies on the possibility to carry out the integration 
using only any observed realisations of the angular distribution $\mathrm{d}N/\mathrm{d}\Omega$
and the event time distribution $\mathrm{d}N/\mathrm{d}T$~:
\begin{equation}
\label{eqn:dist1}
\frac{\mathrm{d}N}{\mathrm{d}\Omega}=\sum_{i=1}^N \delta(\Omega,\Omega_i)=\sum_{i=1}^N \delta(\cos{\theta},\cos{\theta_i})\delta(\varphi,\varphi_i),
\end{equation}
\begin{equation}
\label{eqn:dist2}
\frac{\mathrm{d}N}{\mathrm{d}T}=\sum_{i=1}^N \delta(T,T_i).
\end{equation}
Inserting equation~\ref{eqn:dist1} and equation~\ref{eqn:dist2} into equation~\ref{eqn:shuff1}, it
is straightforward to see that the function $\omega_{sh}(\alpha)$ can be sampled by Monte-Carlo
in \textit{shuffling} the observed set of measured times and in preserving the the
actual set of arrival directions in local coordinates~:
\begin{equation}
\label{eqn:shuff3}
\omega_{sh}(\alpha)\simeq \frac{1}{N_{sh}}\sum_{j=1}^{N_{sh}} \sum_{i=1}^N ~\delta(\alpha(\theta_i,\varphi_i,T_{\sigma_j(i)}),\alpha),
\end{equation}
where the subscript $\sigma_j(i)$ stands for the random permutation of each element $i$.
This estimation is \textit{only} based on any actual data set.

\begin{figure}[!t]
  \centering					 
 \includegraphics[width=10cm]{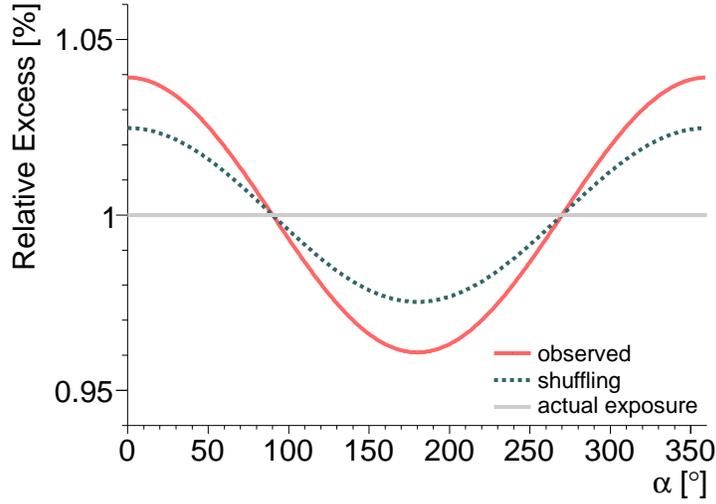}
  \caption{\small{Illustration of the estimation of the directional exposure by means of the shuffling technique
  (in dotted), compared to the true directional exposure (uniform) and to the observed anisotropy (red plain
  curve). For convenience, all curves are here re-scaled around 1.}}
  \label{fig:shuff_principle}
\end{figure}

To illustrate the differences between the actual directional exposure $\omega(\alpha)$ 
(\textit{cf} equation~\ref{eqn:exposure}) and $\omega_{sh}(\alpha)$ as estimated from 
equation~\ref{eqn:shuff1}, let's consider a surface array experiment located at an Earth 
latitude $\ell$, and operating with no variation in time of the detection efficiency so that 
the actual directional exposure $\omega(\alpha)$ is uniform. For a detection efficiency 
saturated up to some maximal zenith angle $\theta_{\mathrm{max}}$, and in a virtual 
situation where an \textit{infinite} number of events would be detected, a dipolar 
distribution with amplitude $d$, declination $\delta_d$ and right ascension $\alpha_d$
would modulate the distributions in local angles very little compared to the case of
isotropy~\cite{AugerApJS2012}, so that 
$\mathrm{d}N(\theta,\varphi)/\mathrm{d}\Omega\propto \cos{\theta}$. On the 
other hand, expressing $\tilde{\Phi}(\delta,\alpha)$ in terms of local sidereal times and 
local angles $\{\theta,\varphi\}$, and integrating over $\theta$ and $\varphi$ leads to~:
\begin{equation}
\label{eqn:shuff4}
\frac{\mathrm{d}N(T)}{\mathrm{d}T}\propto 1+\frac{\displaystyle\int \mathrm{d}\Omega~\cos^2{\theta}}{\displaystyle\int \mathrm{d}\Omega~\cos{\theta}}~d~\cos{\delta_d}~\cos{\ell}\cos{(T-\alpha_d)}.
\end{equation}
Meanwhile, still for a pure dipolar distribution, the distribution in right ascension is known 
from equations~\ref{eqn:nu}~:
\begin{equation}
\label{eqn:shuff4}
\frac{\mathrm{d}N(\alpha)}{\mathrm{d}\alpha}\propto \left(1+\frac{\left<\cos{\delta}\right>d_\perp}{1+\left<\sin{\delta}\right>d_\parallel}~\cos{(\alpha-\alpha_d)} \right),
\end{equation}
where $d_\parallel=d\sin{\delta_d}$ denotes the component of the dipole along the Earth rotation 
axis while $d_\perp=d\cos{\delta_d}$ is the component in the equatorial plane.
For definiteness, all simulated dipoles are oriented in the equatorial plane in the following. 
For $\ell=-35^\circ$ (as in the case of the Pierre Auger observatory~\cite{AugerNIM2004}), 
$d=5\%$, $\delta_d=0^\circ$ and $\alpha_d=0^\circ$ for instance, the function 
$\mathrm{d}N/\mathrm{d}\alpha$ is shown in Figure~\ref{fig:shuff_principle} 
as the red plain curve. Inserting equation~\ref{eqn:shuff3} into equation~\ref{eqn:shuff1}, the function 
$\omega_{sh}(\alpha)$ is shown as the dotted line. 

It is visible from Figure~\ref{fig:shuff_principle} that in contrast to $\omega(\alpha)$, $\omega_{sh}(\alpha)$ 
is not a uniform function~: it absorbs part of the modulation introduced in 
$\mathrm{d}N/\mathrm{d}\alpha$. The anisotropy amplitude estimated by
using $\omega_{sh}(\alpha)$ is consequently \textit{reduced} compared to the one obtained
when using the actual directional exposure function $\omega(\alpha)$. This is the reason why
the sensitivity to large scale anisotropy searches is expected to be reduced when using
the shuffling technique for estimating the directional exposure. 

However, due to fluctuation effects related to the \textit{finite} number of sampled points, the first harmonic 
of \textit{any} observed set of $N$ discrete arrival directions is expected to follow equation~\ref{eqn:shuff4} 
\textit{even for isotropy}, with a random phase and an amplitude distributed according to the Rayleigh 
distribution with parameter $\sigma=\sqrt{2/N}$. Consequently, the non-zero amplitudes resulting from 
fluctuations effects in the case of isotropy are \textit{also} expected to be reduced to some extent. Formally, 
those amplitudes are still expected to follow the Rayleigh distribution but with an effective parameter 
$\sigma_\mathrm{sh}=k\sigma$, with $k$ smaller than 1. This is a crucial mechanism to consider, especially 
for estimating the significance of any measured amplitude by means of the shuffling technique.

Determining the parameter $k$ in a formal semi-analytical way turns out to be a difficult task. This is because 
once the fluctuations associated to the finite number of events are considered, any sets of $a_{\ell \pm1}$ 
coefficients leading to the same first harmonic for the distribution $\mathrm{d}N/\mathrm{d}\alpha$ (\textit{cf}
sub-section~\ref{note}) impact in a different way $\mathrm{d}N/\mathrm{d}T$. In other words, there is no 
one-to-one relation between the first harmonic coefficients of $\mathrm{d}N/\mathrm{d}\alpha$ and the ones
of $\omega_{sh}(\alpha)$. Hence, we defer to Monte-Carlo simulations to calculate the parameter $k$ in 
the following, based on a more sensible toy observatory.

\subsection{Test of the shuffling technique~: a toy observatory}
\label{toyobservatory}

\begin{figure}[!t]
  \centering					 
 \includegraphics[width=7.5cm]{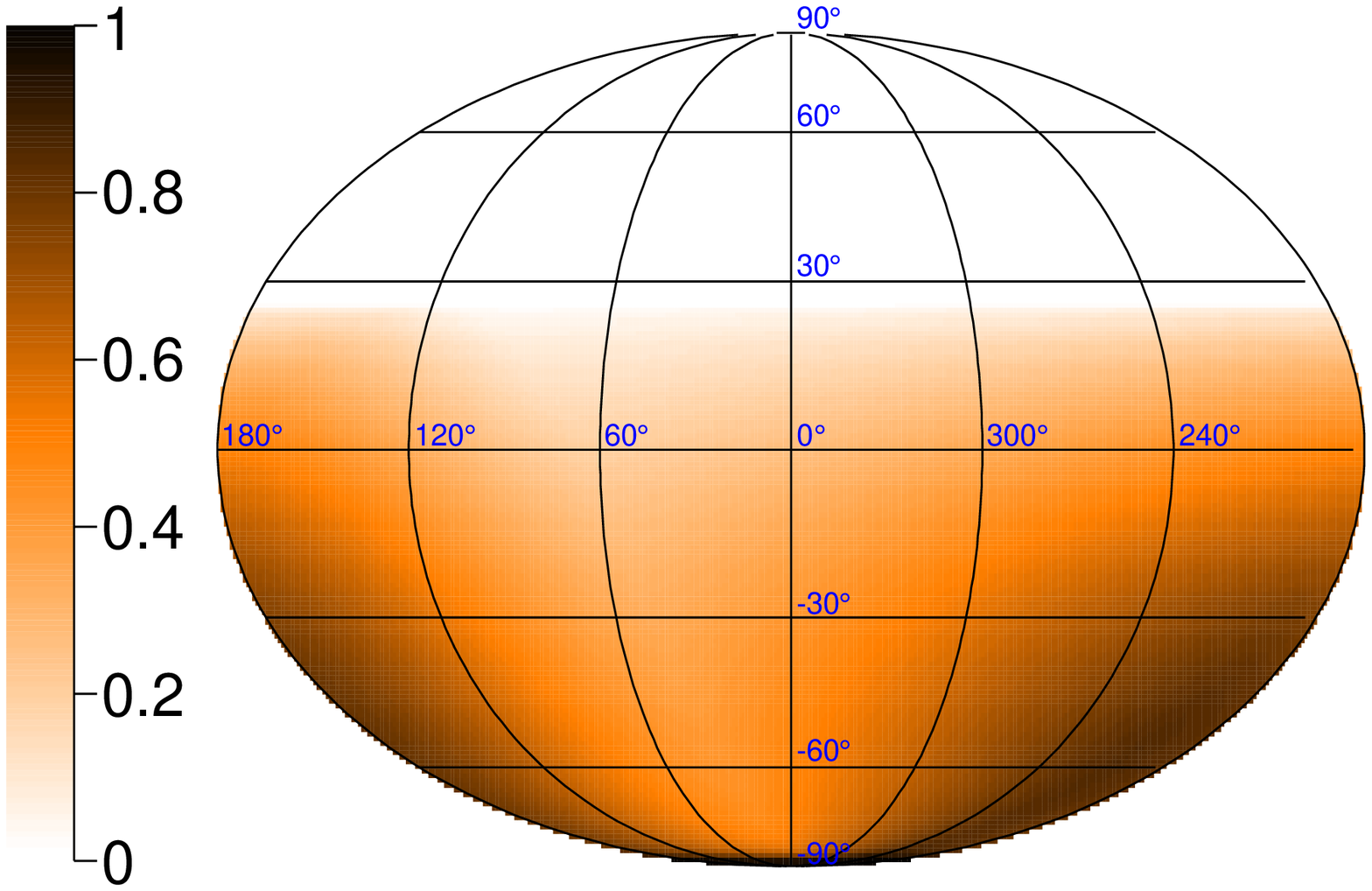}
 \includegraphics[width=7.5cm]{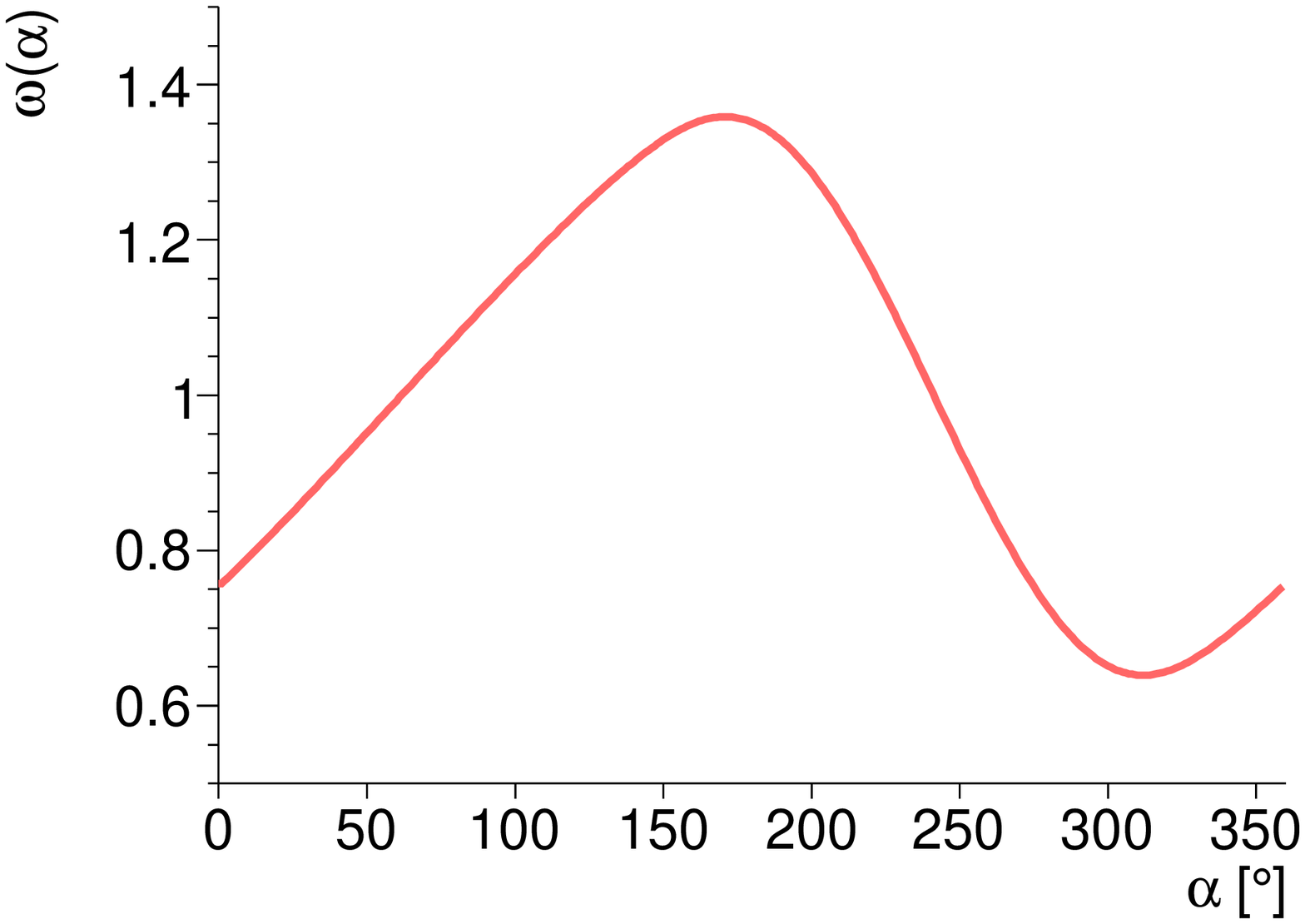}
  \caption{\small{Directional exposure of the toy observatory considered in the Monte-Carlo simulations. 
  Left~: dependence on both right ascension and declination, in Mollweide projection. Right~: dependence
  in right ascension only.}}
  \label{fig:toyexposure}
\end{figure}

To probe the performances of the shuffling technique in searching for large scale anisotropies, 
we will consider in each following sub-section 1,000 mock samples\footnote{In all simulations, 
without loss of generalities, the observatory is assumed to be fully efficient for zenith angles up 
to $60^\circ$.} generated from an isotropic or a dipolar distribution with a total number of events 
$N=100,000$. The toy observatory considered hereafter is assumed to have a detection efficiency
depending on UTC time in such a way that the simulated observatory is operating only 8 hours per 
solar day, with a seasonal harmonic modulation reaching 4 hours. At the sidereal time scale, this 
results in a net dipolar modulation of $\simeq 30\%$ of the directional exposure in right ascension,
as can be seen in Figure~\ref{fig:toyexposure}. The exposure shown in the right panel will be
referred to as the \textit{actual} exposure in the following. Such a behaviour roughly mimics 
fluorescence or Cherenkov telescopes.

Without loss of generalities, we restrict ourselves in the following to the \textit{first} harmonic 
estimation from a set of $N$ events. This is the most challenging to extract from any data set 
since experimental effects mainly produce a spurious dipolar modulation, and one of the most 
interesting for cosmic ray physics at all energies.
 
\subsection{Case of isotropy}
\label{isotropy}

\begin{figure}[!t]
  \centering					 
 \includegraphics[width=7.5cm]{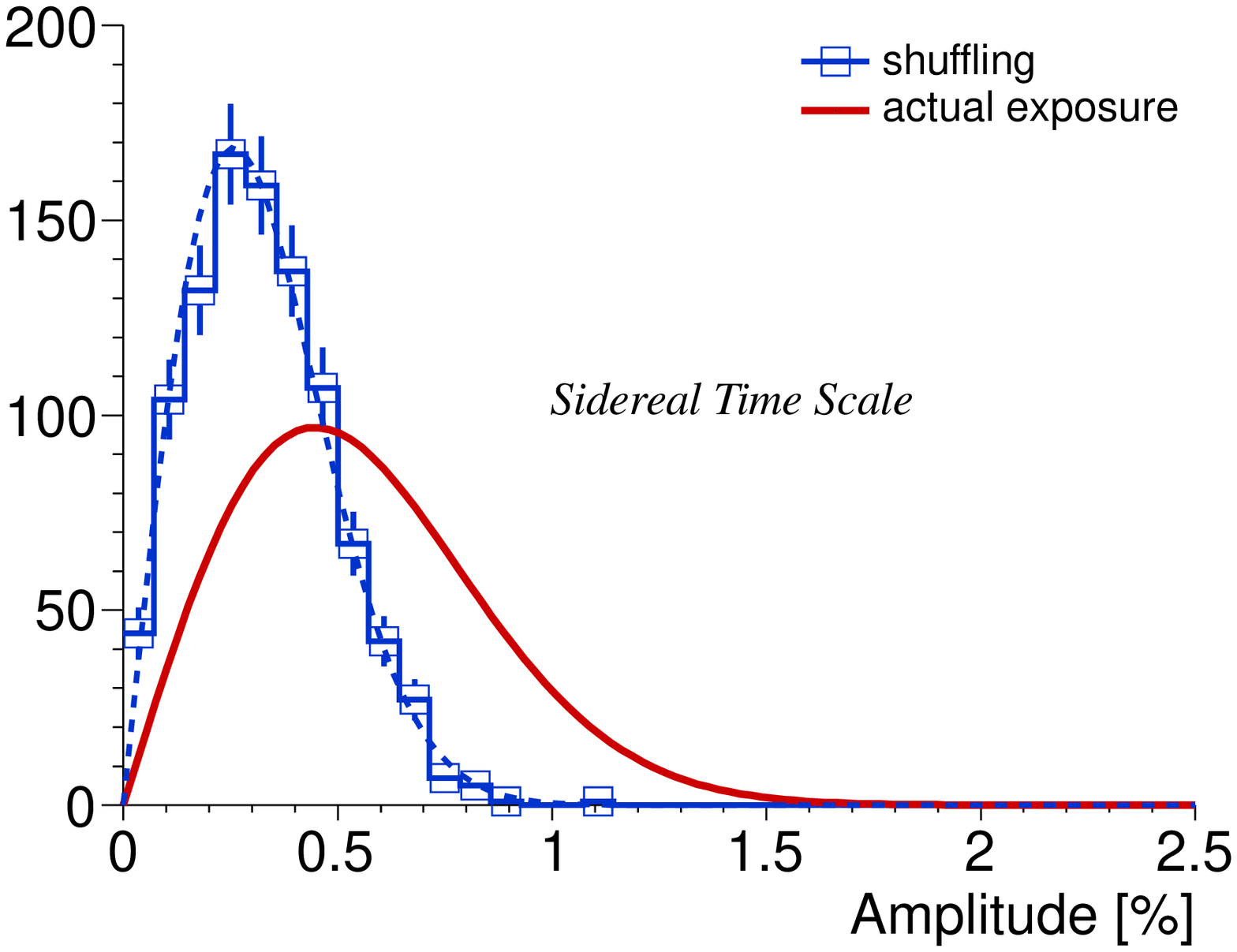}
 \includegraphics[width=7.5cm]{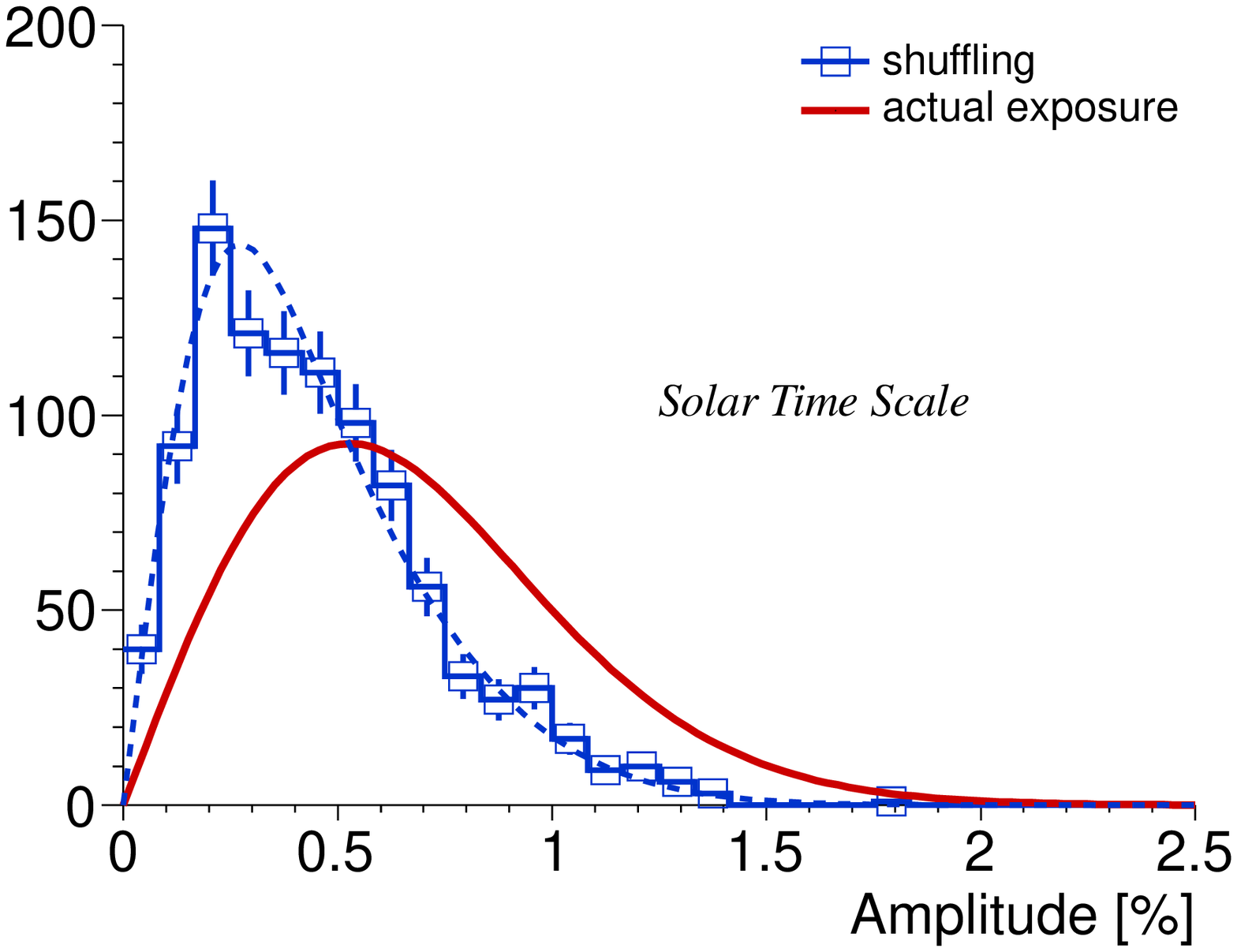}
  \caption{\small{Reconstructed amplitudes for an underlying isotropic distribution, with $N$=100,000 events.
  Plain curves~: Expected distribution of amplitudes from the actual directional exposure. Histograms~: Distribution 
  of reconstructed amplitudes with the directional exposure as derived from the shuffling technique. Left panel~: 
  sidereal time scale. Right panel~: solar time scale.}}
  \label{fig:distiso}
\end{figure}

Generating mock samples drawn from isotropy, the distribution of amplitudes obtained by applying the shuffling 
procedure \textit{on each sample} to estimate the directional exposure is shown in the left panel of 
Figure~\ref{fig:distiso}, as the blue histogram. For reference, the distribution that the histogram of reconstructed 
amplitudes would fit if the actual directional exposure was used is shown as the red curve. Rigorously, this
distribution is described in equation~\ref{eqn:pdf-r}. However, though the directional exposure is largely
non-uniform, the parameters $\sigma_1^c$ and $\sigma_1^s$ differ from $\sigma=\sqrt{2/N}$ only slightly~:
$\sigma_1^c\simeq1.02~\sigma$ and $\sigma_1^c\simeq1.03~\sigma$, so that equation~\ref{eqn:pdf-r}
is here undistinguishable from a Rayleigh distribution with parameter $\sigma$. As a consequence of the 
absorption described above, a \textit{compression} of the reconstructed amplitudes is clearly observed. The 
histogram can still be fitted by the Rayleigh distribution, but by rescaling the $\sigma$ parameter by 
a global factor~: $\sigma_{\mathrm{sh}}\simeq0.57\sigma$. The compression is an essential feature to consider 
when searching for anisotropies, that will be exploited in next sections.

\begin{figure}[!t]
  \centering					 
 \includegraphics[width=7.5cm]{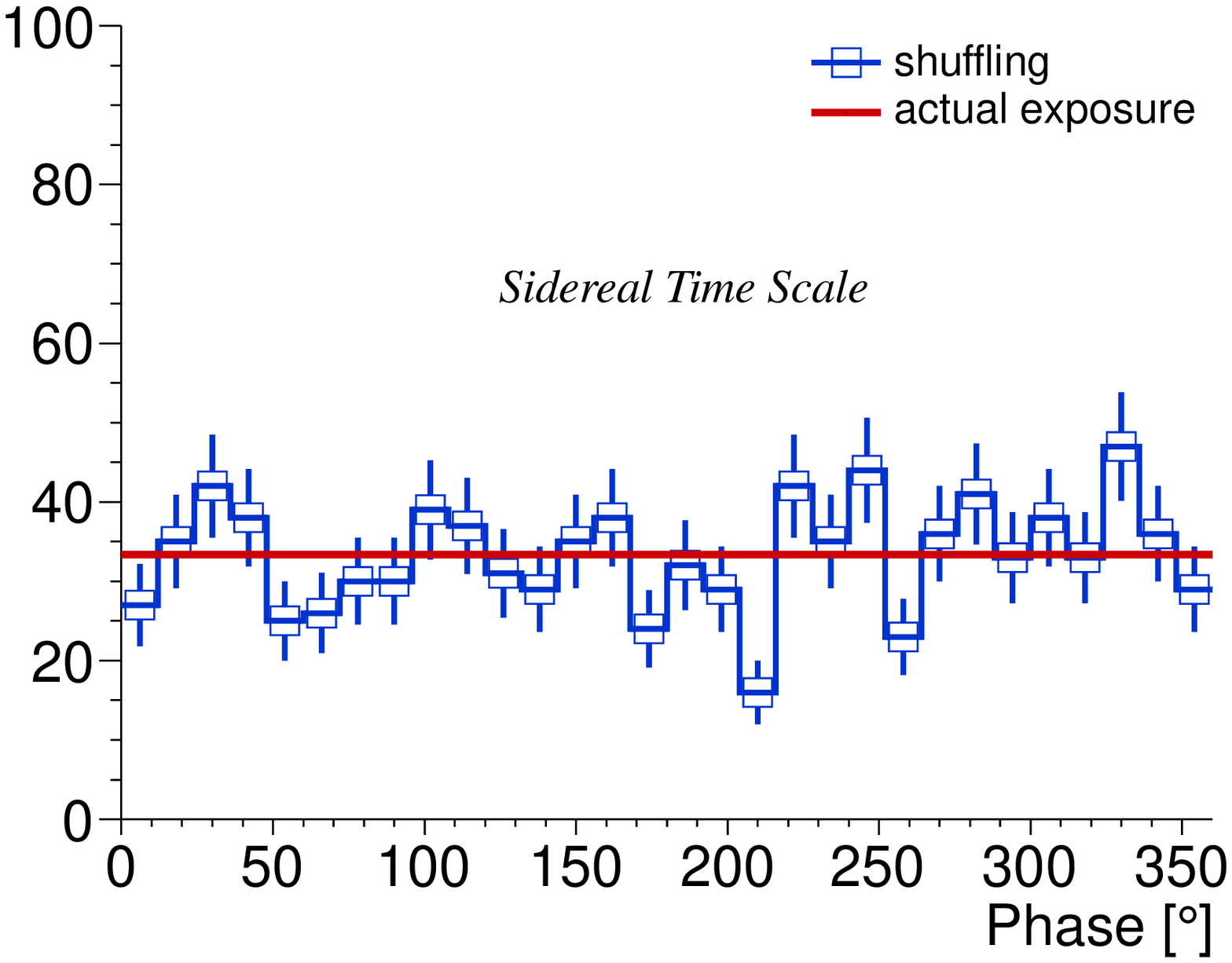}
 \includegraphics[width=7.5cm]{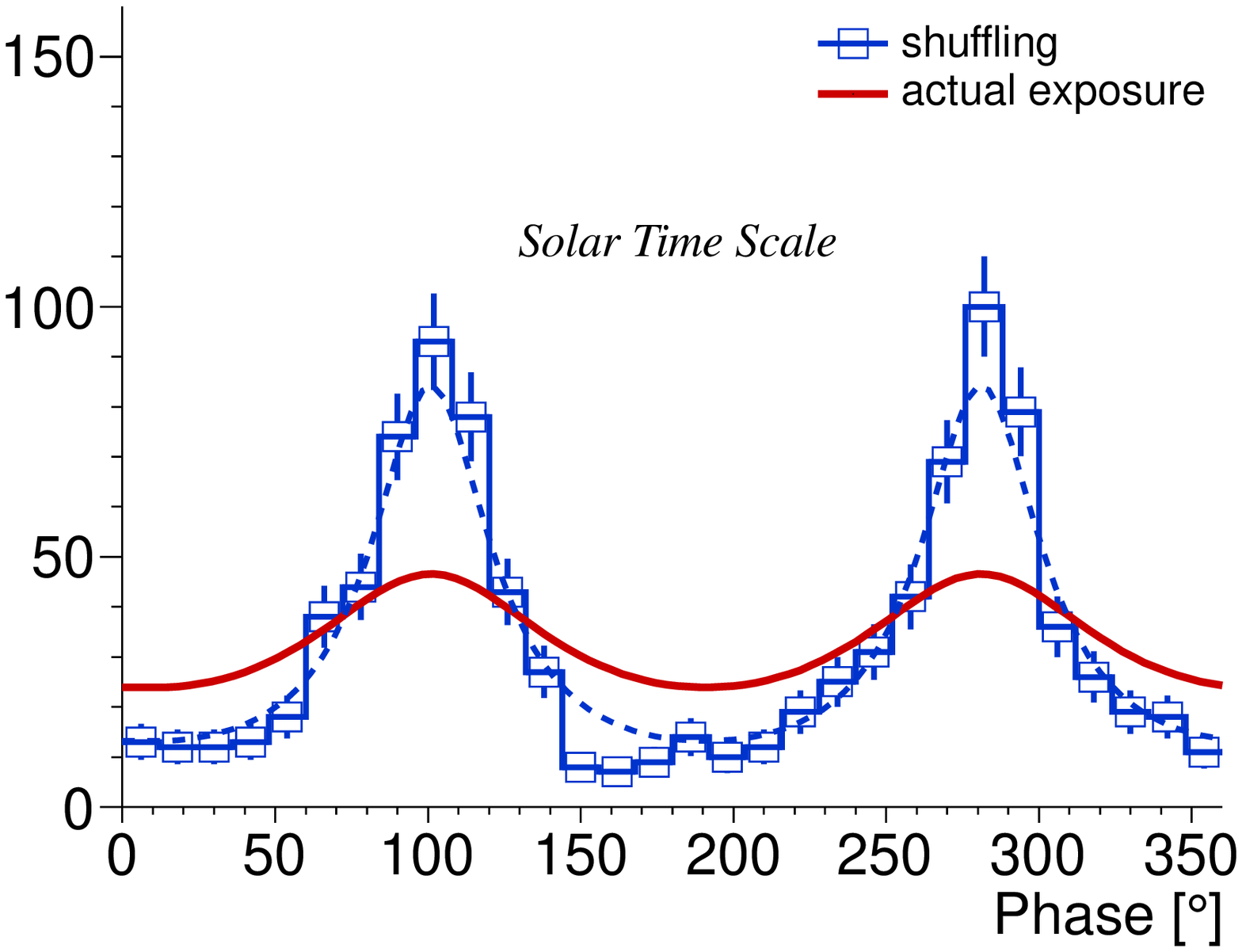}
  \caption{\small{Reconstructed phases for an underlying isotropic distribution, with $N$=100,000 events.
  Plain curves~: Expected distribution of phases from the actual directional exposure. Histograms~: Distribution 
  of reconstructed phases with the directional exposure as derived from the shuffling technique. Left panel~: 
  sidereal time scale. Right panel~: solar time scale.}}
  \label{fig:phaseiso}
\end{figure}

To provide further illustration, the same simulations are repeated at the solar time scale. Given the
time dependence of the efficiency, the net dipolar modulation of the directional exposure is amplified
at this particular time scale, reaching almost $100\%$. This is a \textit{very} significant spurious modulation, 
requiring us to perform here first harmonic analyses using the formalism presented in sub-section~\ref{nonuniform}.
The distribution of amplitudes that would be obtained with the actual directional exposure, shown as the 
red curve in the right panel of Figure~\ref{fig:distiso}, is derived from equation~\ref{eqn:pdf-r} with resolution 
parameters obtained from equation~\ref{eqn:resolution}~: $\sigma_1^c=1.01\sqrt{2/N}$ and 
$\sigma_1^s=1.41\sqrt{2/N}$. In contrast, the distribution of amplitudes obtained by applying the shuffling 
procedure is shown as the blue histogram. A compression is also observed. As in the previous case,
parameters $\sigma_{1,\mathrm{sh}}^c$ and $\sigma_{1,\mathrm{sh}}^s$ can be empirically determined 
so that this histogram can be accurately fitted by equation~\ref{eqn:pdf-r}~: 
$\sigma_{1,\mathrm{sh}}^c\simeq0.42\sigma_1^c$ and $\sigma_{1,\mathrm{sh}}^s\simeq0.77\sigma_1^s$.

Together with the amplitude, the reconstruction of the phase of the first harmonic is a fundamental piece of 
information. As outlined above, the parameters $\sigma_1^c$ and $\sigma_1^s$ (and consequently 
$\sigma_{1,\mathrm{sh}}^c$ and $\sigma_{1,\mathrm{sh}}^s$) are, at the sidereal time scale, too close each 
other to induce notable differences between equation~\ref{eqn:pdf-phi} and a uniform function. The distribution 
of phases is thus expected to be uniform to a high level even when using the shuffling technique. This is observed 
to be the case, as shown in the left panel of Figure~\ref{fig:phaseiso}. On the other hand, at the solar time scale, the 
distribution of phases is expected to follow equation~\ref{eqn:pdf-phi}, with the same $\sigma_{1,\mathrm{sh}}^c$ 
and $\sigma_{1,\mathrm{sh}}^s$ parameters as the ones derived empirically for the distribution of amplitudes. This 
is indeed the case, as shown in the right panel of Figure~\ref{fig:phaseiso} by the dotted curve matching the histogram. 
The distribution that would be obtained using the actual directional exposure is shown as the continuous curve. This 
latter curve is flatter and so closer to a uniform distribution, which is a first illustration of the loss of sensitivity when 
using the shuffling technique - though in an extreme case. This loss of sensitivity will be exemplified and quantified farther. 

\subsection{Case of anisotropy}
\label{anisotropy}

\begin{figure}[!t]
  \centering					 
 \includegraphics[width=10cm]{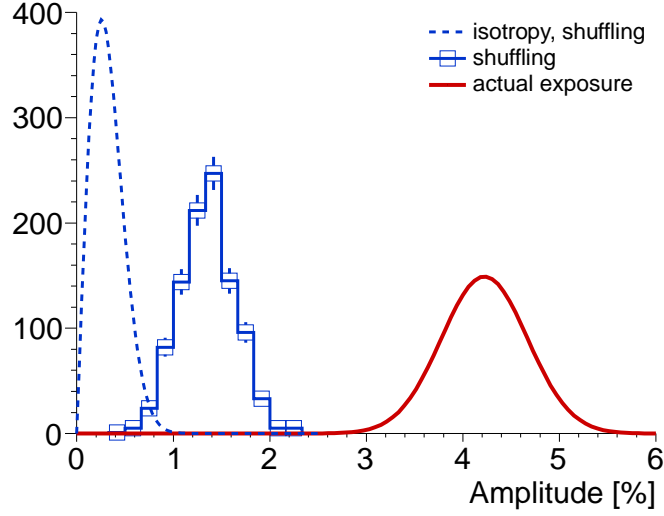}
  \caption{\small{Reconstructed amplitudes for an underlying dipolar distribution with amplitude $d_\perp=5\%$ and 
  $N$=100,000 events. Histogram~: Distribution of reconstructed amplitudes with the directional exposure as derived 
  from the shuffling technique. Plain curve~: Expected distribution of amplitudes when using the true directional exposure.
  Dotted curve~: Distribution of amplitudes for isotropy, as empirically reconstructed from the shuffled event sets built 
  with the directional exposure as derived from each simulated sample.}}
  \label{fig:distaniso}
\end{figure}

We consider now dipolar distributions of cosmic rays, at the sidereal time scale only. By analogy with the isotropic case,
there are no notable differences between equations \ref{eqn:pdf-rice}-\ref{eqn:pdf-psi} and equations~\ref{eqn:pdfanis}
due to the closeness between $\sigma_1^c$ and $\sigma_1^s$ so that, for simplicity, we use 
equations~\ref{eqn:pdf-rice}-\ref{eqn:pdf-psi} in the following discussion, with $\sigma=\sqrt{2/N}$. For a dipole amplitude 
$d_\perp=5\%$, and with the actual directional exposure, the parameters $\nu$ and $\psi$ entering into 
equations~\ref{eqn:pdf-rice}-\ref{eqn:pdf-psi} are given from the set of coefficients in equation~\ref{eqn:nu2}. Their variation 
with the actual dipole phase $\alpha_d$ is almost the same as the illustration depicted in Figure~\ref{fig:nupsi} for the 
observatory latitude at $\pm30^\circ$. Given the toy observatory considered in the simulations, for $d_\perp=5\%$
and $\alpha_d=0^\circ$ for instance, the expectation for $\nu$ is $\simeq 4.2\%$~\footnote{We have checked that this value
for $\nu$ (and the corresponding one for $\psi$) fits the distribution of amplitudes obtained from Monte-Carlo simulations.}. 
The corresponding distribution of 
reconstructed amplitudes is shown as the plain red curve in Figure~\ref{fig:distaniso}. On the other hand, the distribution 
of reconstructed amplitudes with the directional exposure as derived from the shuffling technique is shown as the blue 
histogram. This histogram can be fitted by equation~\ref{eqn:pdf-rice} by rescaling the $\sigma$ parameter in the same 
way as in the case of isotropy (\textit{i.e.} $\sigma_{\mathrm{sh}}\simeq0.57\sigma$) and with a $\nu$ parameter such that 
$\nu_\mathrm{sh}=1.32\%\pm0.01\%$ instead of $\nu=4.2\%$. Though part of the anisotropy signal is, as expected, absorbed, 
it can be seen however that significant amplitudes can be discerned by comparison of the blue histogram in 
Figure~\ref{fig:distaniso} with the one in Figure~\ref{fig:distiso}.

\begin{figure}[!t]
  \centering					 
 \includegraphics[width=10cm]{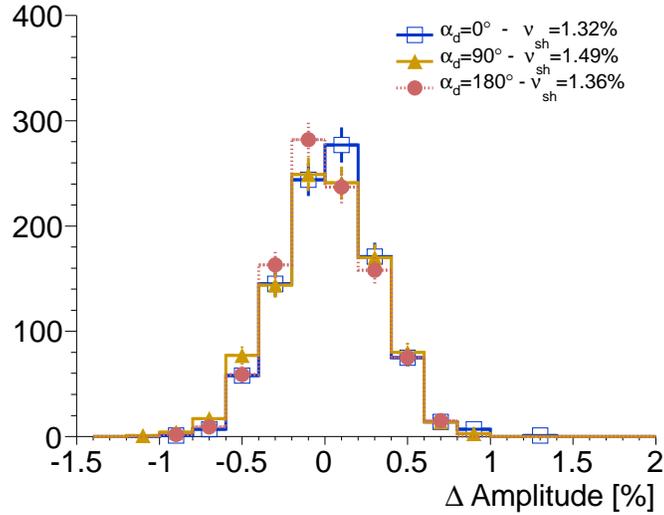}
  \caption{\small{Reconstructed amplitudes centered on the expected $\nu_\mathrm{sh}$ parameter for an underlying dipolar 
  distribution with amplitude $d_\perp=5\%$, $N$=100,000 events, and different dipole phases $\alpha_d$. The directional 
  exposure is derived from the shuffling technique.}}
  \label{fig:rnu}
\end{figure}

Though this is irrelevant in practice since $\tilde{\omega}(\delta,\alpha)$ is by construction not accessible to knowledge 
when using the shuffling technique, it is anyway interesting to check here that the resulting value for $\nu_\mathrm{sh}$ 
is in agreement with the one that can be derived from equations~\ref{eqn:nu2} when inserting the numerical values
of $\omega_\mathrm{sh}$ averaged over many realisations in the denominators of these equations. This turns out 
to be the case, as illustrated in Figure~\ref{fig:rnu}, where the expected $\nu_\mathrm{sh}$ is subtracted to the 
reconstructed amplitudes for several values of dipole phases $\alpha_d$. All histograms can be observed to be
centered on zero. 

Hence, from any single mock sample, it appears interesting to aim at reconstructing in an empirical way
a relevant distribution of amplitudes that would have been obtained with an underlying isotropic angular
distribution of cosmic rays. This can be done using the following recipe~:
\begin{enumerate}
\item From any given data set, the directional exposure is obtained by means of the shuffling technique, 
allowing an amplitude $\overline{r}_1^{obs}$ to be inferred.
\item New mock samples are then generated, still by means of the shuffling technique.
\item The shuffling procedure is applied to each mock sample generated in step 2, allowing
the directional exposure of each mock sample to be known. The corresponding distribution of amplitudes 
$\overline{r}_1$ is then constructed in an empirical way.
\item The amplitude $\overline{r}_1^{obs}$ is then compared to the empirical distribution obtained in step 3.
\end{enumerate}
The result of this procedure is shown in Figure~\ref{fig:distaniso} too, as the blue \textit{dotted} curve. Here, for clarity,
what is shown is not the histogram obtained for any realisation but the corresponding fit using equation~\ref{eqn:pdf-r}.
It turns out that the distribution of amplitudes obtained in that way is in agreement with the one obtained
in previous sub-section~\ref{isotropy} \textit{for isotropy}. In other words, the procedure described above provides a 
distribution of amplitudes that makes it possible to estimate the significance of any measured amplitude 
obtained with a directional exposure estimated by means of the shuffling technique. \textit{This procedure 
is only based on any given data set.}

\begin{figure}[!t]
  \centering					 
 \includegraphics[width=10cm]{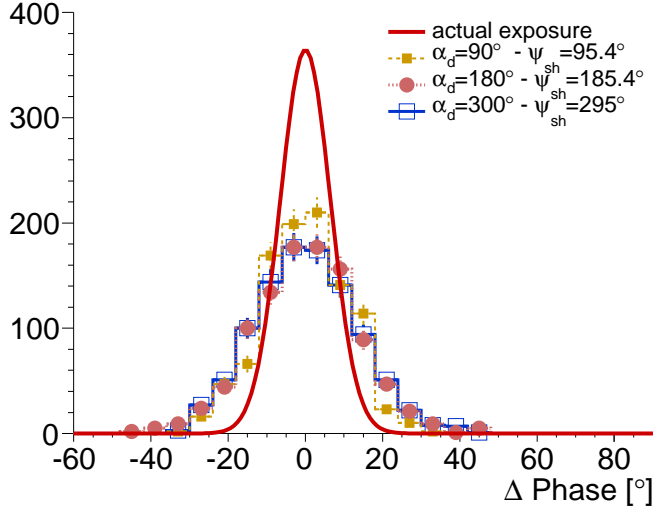}
  \caption{\small{Reconstructed phases for an underlying dipolar distribution with amplitude $d_\perp=5\%$ and 
  $N$=100,000 events. Histograms~: Distribution of reconstructed phases with the directional exposure as derived 
  from the shuffling technique, and for three directions of the dipole phase. Plain curve~: Expected distribution of 
  phases when using the actual directional exposure.}}
  \label{fig:phaseaniso}
\end{figure}

For completeness, the distributions of reconstructed phases are shown as the histograms in Figure~\ref{fig:phaseaniso},
\textit{after subtraction of the expected $\psi$ and $\psi_\mathrm{sh}$} for both the actual exposure $\omega(\alpha)$ 
and the shuffling one $\omega_\mathrm{sh}(\alpha)$ respectively. The distribution that would be obtained using the actual 
directional exposure is shown again as the red plain curve. On the other hand, the reconstructed distributions using the 
shuffling technique are shown as the blue histograms for three directions of the dipole phase~: one along the direction of 
maximum of exposure $\alpha_d=180^\circ$, a second one in the direction of minimum of exposure $\alpha_d=300^\circ$, 
and a third one in the direction $\alpha_d=90^\circ$. All distributions are centered and can be fitted by equation~\ref{eqn:pdf-psi} 
with the same parameters $\sigma_\mathrm{sh}$ and $\nu_\mathrm{sh}$ as the ones derived from Figure~\ref{fig:distaniso}. 
The resulting width of these curves is larger than the one that would be obtained in the ideal case, which is a new illustration 
of the loss of sensitivity when using the shuffling technique, loss of sensitivity that we quantify in next section. 

It is important to stress that, in the light of results obtained in section~\ref{anisotropiccases}, the parameters $\nu_\mathrm{sh}$ 
and $\psi_\mathrm{sh}$ are related to the dipole parameters $d_\perp$ and $\alpha_d$
in an inaccessible way in the case of a non-uniform directional exposure in right ascension since the directional exposure 
function $\tilde{\omega}(\delta,\alpha)$ is supposed to be unknown in the practical cases where the shuffling technique is relevant. 
In practice, the only possible way to convert $\nu_\mathrm{sh}$ and $\psi_\mathrm{sh}$ into $d_\perp$ and $\alpha_d$ thus requires 
the use of equation~\ref{eqn:nu} (that is, ignoring the non-uniformities of the directional exposure in right ascension). This approximation 
leads unavoidably to small biases for the estimation of $d_\perp$ and $\alpha_d$.

\section{Performances of the shuffling technique for large scale anisotropy searches}
\label{sec:performances}

\subsection{Detection power}
The performances of the procedure described in previous section in searching for real anisotropies
are presented in this section. They are given in terms of \textit{detection power} at some \textit{threshold} value.
The threshold - or type-I error rate - is the fraction of isotropic simulations in which the null hypothesis is wrongly
rejected (\textit{i.e.} the test provides evidence of anisotropy when there is no anisotropy). On the other hand, the
detection power, defined as $1-\beta$ with $\beta$ the type-II error rate, provides the fraction of anisotropic
simulations in which the null hypothesis is (correctly) rejected. 

\begin{figure}[!t]
  \centering					 
 \includegraphics[width=10cm]{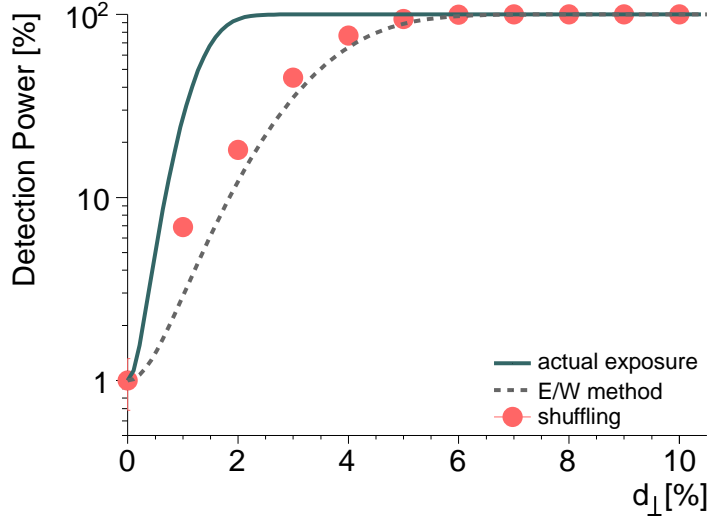}
  \caption{\small{Detection power for dipolar anisotropy searches as a function of $d_\perp$. The threshold is fixed
  at 1\%. The results obtained using the shuffling (red points) are compared to the ones obtained with the actual exposure
  (plain curve) and with the East/West method (dotted curve).}}
  \label{fig:power}
\end{figure}

Still with mock samples with a total number of events $N=100,000$ and in the same conditions of detection 
efficiency as in sub-section~\ref{anisotropy}, the detection power is shown in Figure~\ref{fig:power} as a function 
of $d_\perp$, and for random phases $\alpha_d$. The threshold is fixed here at 1\%. The results obtained using 
the shuffling (red points) are compared to the ones obtained with the actual exposure (plain curve).
For illustration, the results obtained using the East/West method~\cite{Bonino2011} are also shown as the 
dotted curve. This method is based on the analysis of the difference of the counting rates in the eastward
and the westward directions, and is largely independent of experimental effects without requiring corrections
for directional exposure and/or atmospheric effects. To maximise the performances of the East/West method,
our simulated observatory is assumed to present a perfect symmetry to eastward and westward directions.
Though the detection power is saturated much earlier in the case of the first harmonic analysis using the 
actual directional exposure, it turns out that the procedure described in sub-section~\ref{anisotropy} performs 
slightly better than the East/West method. In addition, it is worth noting that the procedure presented here 
potentially pertains to any type of detector (\textit{i.e.} even for fluorescence or Cherenkov telescopes).

\subsection{From the measured amplitude to $d_\perp$}

\begin{figure}[!t]
  \centering					 
 \includegraphics[width=10cm]{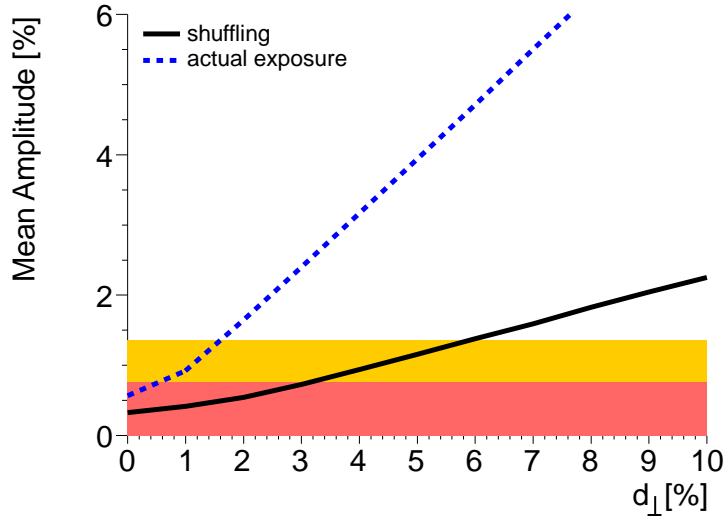}
  \caption{\small{Mean amplitude of the first harmonic as a function of $d_{\perp}$.}}
  \label{fig:rmeanvsd}
\end{figure}

At this stage, the procedure presented above provides a relevant framework to reveal large scale
anisotropy. However, for cosmic ray physics, it is interesting not only to detect anisotropy but also
to estimate $d_\perp$. This task is straightforward to achieve when the actual directional
exposure is known and when a pure dipole is assumed by means of equation~\ref{eqn:shuff4}. This is 
illustrated by the dotted blue line in Figure~\ref{fig:rmeanvsd}. For convenience, the amplitude range that 
can be obtained in 99\% of isotropic fluctuations is delimited by the orange area. 

In contrast, when using the shuffling technique to get at the exposure, equation~\ref{eqn:shuff4} does not
hold any longer. To estimate the conversion curve between
$\left<r\right>$ and $d_\perp$, the procedure described in sub-section~\ref{anisotropy} can be 
applied by replacing the second step in this way~: \textit{New mock samples are then generated, 
still by means of the shuffling technique. In addition, each event is accepted or rejected by 
randomly sampling its right ascension according to the modulation in right ascension
induced by a dipole with amplitude $d_\perp$ and phase $\alpha_d$}. The resulting curve obtained
in the same simulated experimental conditions as previously is shown as the black curve in 
Figure~\ref{fig:rmeanvsd}, together with the amplitude range that can be obtained in 99\% of isotropic 
fluctuations (red area). This curve allows a \textit{calibration} of the measured amplitude in terms 
of $d_\perp$. For instance, a measured amplitude $\overline{r}^{obs}=1\%$, providing
evidence for anisotropy at more than 99\% confidence level, corresponds in fact to $d_\perp=4\%$
in this specific example.

\section{Conclusions}
\label{conclusions}

The shuffling technique has been studied to search for large scale anisotropies of cosmic rays.
Accounting for the compression of the reconstructed amplitude distribution in the case of an underlying
isotropic distribution, it has been shown that this technique can be used to reveal dipolar patterns
though with reduced sensitivity compared to the one that would be obtained with the perfect 
knowledge of the actual directional exposure. In spite of this reduced sensitivity, the gain of this 
method relies on avoiding any corrections for the observed counting rates. This holds for any 
surface array detectors, fluorescence telescopes, and/or Cherenkov telescopes.

\section*{Acknowledgements}
We thank Piera Ghia for her careful reading of the text and helpful suggestions, and the referee of
Astroparticle Physics whose comments helped us to improve the paper.


\end{document}